\documentclass[11pt,reqno]{amsart}
\usepackage{amsmath,amsfonts,amscd,amssymb,epsf}
\newtheorem{theorem}{Theorem}[section]
\newtheorem{proposition}[theorem]{Proposition}
\newtheorem{lemma}[theorem]{Lemma}

\numberwithin{equation}{section}

\newcommand{\pa}{\partial}

\newcommand{\bs}{\boldsymbol}

\begin{document}
\title {The Multicomponent KP Hierarchy:   Differential Fay Identities and Lax Equations}
\author{Lee Peng Teo}\address{Department of Applied Mathematics, Faculty of Engineering, University
of Nottingham Malaysia Campus, Jalan Broga, 43500, Semenyih, Selangor
Darul Ehsan, Malysia. }\email{LeePeng.Teo@nottingham.edu.my  }

\begin{abstract}
In this article, we show that four sets of differential Fay identities of an $N$-component KP hierarchy derived from the bilinear relation satisfied by the tau function of the hierarchy are sufficient to derive the auxiliary linear equations for the wave functions. From this, we derive the Lax representation for the $N$-component KP hierarchy, which are equations satisfied by some pseudodifferential operators with matrix coefficients. Besides the Lax equations with respect to the time variables proposed in \cite{2}, we also obtain a set of equations relating different charge sectors, which can be considered as a generalization of the modified KP hierarchy proposed in \cite{3}.
\end{abstract}

\maketitle

\section{Introduction}

The KP hierarchy \cite{6} is one of the most extensively studied integrable hierarchies. It arises in many different fields of mathematics and physics such as enumerative algebraic geometry, hydrodynamics and string theory. It is an infinite set of coupled partial differential equations of infinitely many functions $u_1, u_2,\ldots$ of infinitely many variables $t_1,t_2,\ldots$. In terms of the pseudo-differential operator $L=\pa + u_1\pa^{-1}+u_2\pa^{-2}+\ldots$, where $\pa=\pa_{t_1}$, the partial differential equations can be expressed as
$$\frac{\pa L}{\pa t_n}=[B_n, L],\hspace{1cm} B_n:=(L^n)_+,\quad n=1,2,\ldots,$$ where $(L^n)_+$ means the differential part of the operator $L^n$. One of the biggest breakthroughs in the study of the KP hierarchy is the group theoretical  description of the solutions  of the KP hierarchy \cite{6}, which is closely related to the infinite dimensional Grassmann manifolds \cite{8,7}. To every solution of the KP hierarchy, there exists a tau function $\tau(\boldsymbol{t})$ which satisfies the bilinear relation
\begin{equation}\label{eq10_28_1}\oint  \tau(\bs{t}-[z^{-1}]) e^{\sum_{n=1}^{\infty} (t_n-t_n')z_n}\tau(\bs{t}'+[z^{-1}])dz=0.\end{equation}Here $\bs{t}=(t_1,t_2,\ldots)$ and $[z^{-1}]=(z^{-1}, z^{-2}/2, z^{-3}/3,\ldots)$.
 Such a tau function can be represented using the charge zero sector of a free fermion system.

In \cite{2},   Date,   Jimbo,   Kashiwara and   Miwa extended their work \cite{6} to the multicomponent KP hierarchy proposed by Sato in a lecture. For an $N$-component KP hierarchy, there are $N$ infinite families of time variables $t_{\alpha n}, \alpha=1,\ldots,N, n=1,2,\ldots$. The coefficients $u_1, u_2,\ldots$ of the Lax operator $L=\pa +u_1\pa^{-1}+u_2\pa^{-2}+\ldots$ are $N\times N$ matrices. The operator $\pa$ now is equal to $\pa_{t_{11}}+\ldots +\pa_{t_{N1}}$. There are another $N$ pseudodifferential operators $R_1,\ldots, R_N$ of the form
$$R_{\alpha}=E_{\alpha}+u_{\alpha 1}\pa^{-1}+u_{\alpha 2}\pa^{-2}+\ldots,$$ where $E_{\alpha}$ is the $N\times N$ matrix with $1$ on the $(\alpha,\alpha)$-component and zero elsewhere, and $u_{\alpha 1}, u_{\alpha 2},\ldots$ are also $N\times N$ matrices. The operators $L, R_1,\ldots, R_N$ satisfy the following conditions:
$$LR_{\alpha}=R_{\alpha} L, \quad R_{\alpha}R_{\beta}=\delta_{\alpha\beta}R_{\alpha}, \quad \sum_{\alpha=1}^N R_{\alpha}= \boldsymbol{1}.$$ The Lax equations are:
\begin{equation*}
\frac{\pa L}{\pa t_{\alpha n}}=[B_{\alpha n}, L],\hspace{1cm}\frac{\pa R_{\beta}}{\pa t_{\alpha n}}=[B_{\beta n}, L],\hspace{1cm}B_{\alpha n}:=(L^n R_{\alpha})_+.
\end{equation*}
The tau function of an $N$-component KP hierarchy can also be expressed in terms of fermions, but $N$ components of free fermions are required. Moreover, one has to go beyond the charge zero sector. More precisely, let $s_1, \ldots, s_N$ be the charge of each component of the free fermions. Then for a fixed $\bs{s}=(s_1,\ldots,s_N)^T$ with $s_1+\ldots+s_N=0$, the tau function of an $N$-component KP hierarchy can be written as $\tau(\bs{s}, \bs{t})$, where $\bs{t}$ is the collective notation for all the time variables. $\tau(\bs{s},\bs{t})$ satisfies the following bilinear relations:
\begin{equation} \label{eq10_28_2}
\begin{split}
&\sum_{\gamma=1}^N\epsilon_{\alpha\gamma}(\boldsymbol{s})\epsilon_{\beta\gamma}(\boldsymbol{s}')\oint dz z^{s_{\gamma}-s_{\gamma}'+\delta_{\alpha\gamma}+\delta_{\beta\gamma}-2}e^{\xi(\boldsymbol{t}_{\gamma}-\boldsymbol{t}_{\gamma}',z)}\\&\hspace{1cm}\times
\tau\left(\boldsymbol{s}
+\boldsymbol{e}_{\alpha}-\boldsymbol{e}_{\gamma},\boldsymbol{t}-[z^{-1}]_{\gamma}\right)
\tau\left(\boldsymbol{s}'-\boldsymbol{e}_{\beta}+\boldsymbol{e}_{\gamma}, \boldsymbol{t}'+ [z^{-1}]_{\gamma}\right)=0
\end{split}
\end{equation}for any $1\leq \alpha,\beta\leq N$. Here $\bs{e}_{\alpha}$ is the column vector with $1$ on the $\alpha$-position and zero elsewhere, and $\epsilon_{\alpha\gamma}(\boldsymbol{s})$ is a sign function. One problem that has not been explored in connection with he multicomponent KP hierarchy is the dependence of the operators $L, R_1,\ldots, R_N$ on the charge variables $\bs{s}$. This will be considered in this article.

In the seminal paper \cite{9}, Takasaki and Takebe derived the differential Fay identity for the KP hierarchy from the bilinear relation \eqref{eq10_28_1}. It was shown that the differential Fay identity is equivalent to the KP hierarchy, by first showing that the differential Fay identity implies linear equations of the form
$$\pa_{t_n} \Psi=B_n(\pa)\Psi,$$ where $\Psi$ is the wave function of the KP hierarchy, and $B_n$ is a differential operator in $\pa=\pa_{t_1}$ of order $n$.  In \cite{10}, Takasaki derived the differential Fay identities of BKP and DKP hierarchies and obtained the auxiliary linear equations  of these hierarchies from their respective differential Fay identities. This further illustrates the importance of differential Fay identities as a set of identities that encode all the information of the integrable hierarchies. Differential Fay identities also play    important roles in studying the dispersionless limits of integrable hierarchies. In \cite{1}, Takasaki and Takebe derived   four set of differential Fay identities for the multicomponent KP hierarchy from the bilinear relation \eqref{eq10_28_2} and showed that their dispersionless limits give rise to the universal Whitham hierarchy. In fact, they considered $(N+1)$-component KP hierarchy, where one of the components is more special than the other components and is denoted by the $0^{\text{th}}$-component. Some auxiliary linear equations for $(N+1)$ components of the matrix wave function were derived from the differential Fay identities but they   do not directly lead to the Lax representation of the multicomponent KP hierarchy. In fact, for an $(N+1)$-component KP hierarchy, the wave function is an $(N+1)\times (N+1)$-matrix valued function, but only one row of this matrix, the $0^{\text{th}}$-row, was considered in \cite{1}. The auxiliary linear equations derived in \cite{1} only leads to linear evolution equations of the form
$$\pa_{t_{\alpha n}}\Psi_0=B_{\alpha n}(\pa_{\alpha 1})\Psi_0,$$ where $\Psi_0$ is the $0^{\text{th}}$-row of the matrix wave function, and $B_{\alpha n}(\pa_{\alpha 1})$ is a differential operator of order $n$ in $\pa_{t_{\alpha 1}}$. One would expect that for a multicomponent KP hierarchy, the $B_{\alpha n}$ is a differential operator of $\pa=\pa_{t_{11}}+\ldots+\pa_{t_{N1}}$ rather than of $\pa_{t_{\alpha 1}}$. In this article, we show that without singling out a special component, one can derive the auxiliary linear equations  for the matrix wave function   of the multicomponent KP hierarchy  from the differential Fay identities which have the expected form
$$\pa_{t_{\alpha n}}\Psi=B_{\alpha n}(\pa)\Psi.$$ Besides these linear differential equations with respect to the time variables $t_{\alpha n}$, we also obtain  linear equations with respect to the charge variables. The latter is what we need for exploring the variations of the operators $L, R_1,\ldots, R_N$ with respect to the charge variables $s_1,\ldots, s_N$.

\section{ $N$-component KP hierarchy}
Let $N$ be a positive integer. The time variables of the $N$-component KP hierarchy are $N$-sequences of continuous variables $\{t_{1j}\}_{j=1}^{\infty}$,\ldots, $\{t_{Nj}\}_{j=1}^{\infty}$ collectively denoted by
\begin{equation*}
\boldsymbol{t}=\left(\boldsymbol{t}_1,\boldsymbol{t}_2,\ldots,\boldsymbol{t}_N\right)^T\hspace{1cm}\boldsymbol{t}_{\alpha}=(t_{\alpha1}, t_{\alpha2},\ldots),\hspace{0.5cm}\alpha=1,2,\ldots,N.
\end{equation*}There are  $N$ additional discrete charge variables $s_1, s_2, \ldots, s_N\in \mathbb{Z}$, collectively written as $$\boldsymbol{s}=( s_1,\ldots,s_N)^T,  $$and subjected to the condition$$\sum_{\alpha=1}^{N}s_{\alpha}=0.
$$An $N$-component KP hierarchy can be defined by the bilinear identities satisfied by its tau function $\tau(\boldsymbol{s},\boldsymbol{t})$ \cite{2,4,1,5}:
\begin{equation}\label{eq12_24_1}
\begin{split}
&\sum_{\gamma=1}^N\epsilon_{\alpha\gamma}(\boldsymbol{s})\epsilon_{\beta\gamma}(\boldsymbol{s}')\oint dz z^{s_{\gamma}-s_{\gamma}'+\delta_{\alpha\gamma}+\delta_{\beta\gamma}-2}e^{\xi(\boldsymbol{t}_{\gamma}-\boldsymbol{t}_{\gamma}',z)}\\&\hspace{1cm}\times
\tau\left(\boldsymbol{s}
+\boldsymbol{e}_{\alpha}-\boldsymbol{e}_{\gamma},\boldsymbol{t}-[z^{-1}]_{\gamma}\right)
\tau\left(\boldsymbol{s}'-\boldsymbol{e}_{\beta}+\boldsymbol{e}_{\gamma}, \boldsymbol{t}'+ [z^{-1}]_{\gamma}\right)=0
\end{split}
\end{equation}for any $1\leq \alpha,\beta\leq N$.
Here $\boldsymbol{e}_{\alpha}$ is the $N\times 1$ column vector with $1$ on the $\alpha^{\text{th}}$ place and zero elsewhere,
$$\epsilon_{\alpha\gamma}(\boldsymbol{s})=\begin{cases}
(-1)^{s_{\alpha+1}+\ldots+s_{\gamma}}, \hspace{0.5cm}&\text{if}\;\;\alpha<\beta\\
\hspace{1cm}1, & \text{if}\;\; \alpha=\beta\\
-(-1)^{s_{\gamma+1}+\ldots+s_{\alpha}}, &\text{if}\;\;\alpha>\beta
\end{cases},$$
$$\xi(\boldsymbol{t}_{\gamma},z)=\sum_{j=1}^{\infty} t_{\gamma j}z^j,$$ and $$\left(\boldsymbol{t}-[z^{-1}]_{\gamma}\right)_{\alpha j} =
t_{\alpha j}-\delta_{\alpha\gamma}\frac{z^{-j}}{j}.$$As pointed out in \cite{1}, it is sufficient to consider the case where $\alpha=\beta$, which gives
\begin{equation}\label{eq12_24_2}
\begin{split}
&\oint dz z^{s_{\alpha}-s_{\alpha}' }e^{\xi(\boldsymbol{t}_{\alpha}-\boldsymbol{t}_{\alpha}',z)}\tau\left(\boldsymbol{s},\boldsymbol{t}-[z^{-1}]_{\alpha}\right)
\tau\left(\boldsymbol{s}', \boldsymbol{t}'+ [z^{-1}]_{\alpha}\right) 
 +\sum_{\substack{1\leq \gamma\leq N\\\gamma\neq\alpha}}\epsilon_{\alpha\gamma}(\boldsymbol{s})\epsilon_{\alpha\gamma}(\boldsymbol{s}')\\&\times\oint dz z^{s_{\gamma}-s_{\gamma}'-2}e^{\xi(\boldsymbol{t}_{\gamma}-\boldsymbol{t}_{\gamma}',z)}\tau\left(\boldsymbol{s}
+\boldsymbol{e}_{\alpha}-\boldsymbol{e}_{\gamma},\boldsymbol{t}-[z^{-1}]_{\gamma}\right)
\tau\left(\boldsymbol{s}'-\boldsymbol{e}_{\alpha}+\boldsymbol{e}_{\gamma}, \boldsymbol{t}'+ [z^{-1}]_{\gamma}\right)=0.
\end{split}
\end{equation}The general case \eqref{eq12_24_1} where $\alpha\neq \beta$ can be recovered from \eqref{eq12_24_2} by replacing $\boldsymbol{s}'$ with $\boldsymbol{s}'+\boldsymbol{e}_{\alpha}-\boldsymbol{e}_{\beta}$, using the fact that
\begin{lemma}\label{lemma1}
Let $1\leq \alpha,\beta,\gamma \leq N$ be three distinct integers. For any $\boldsymbol{s}$, we have
\begin{enumerate}
\item[(i)] $\displaystyle \epsilon_{\alpha\beta}(\boldsymbol{s}+\boldsymbol{e}_{\alpha}-\boldsymbol{e}_{\beta})=\epsilon_{\beta\alpha}(\boldsymbol{s})$,
\item[(ii)] $\displaystyle \epsilon_{\alpha\gamma}(\boldsymbol{s}+\boldsymbol{e}_{\alpha}-\boldsymbol{e}_{\beta})= \epsilon_{\beta\gamma}(\boldsymbol{s})
\epsilon_{\beta\alpha}(\boldsymbol{s})$.
\end{enumerate}\end{lemma}

The wave function $\Psi(\boldsymbol{s},\boldsymbol{t},z)$ and the adjoint wave function $\Psi^*(\boldsymbol{s},\boldsymbol{t},z)$ of the $N$-component KP hierarchy are $N\times N$ matrix-valued functions with the $(\alpha,\beta)$-components defined respectively by
\begin{equation}\label{eq12_24_9}\begin{split}\Psi_{\alpha\beta}(\boldsymbol{s},\boldsymbol{t},z)=&\epsilon_{\alpha\beta}(\boldsymbol{s})\frac{\tau(\boldsymbol{s}+\boldsymbol{e}_{\alpha}-\boldsymbol{e}_{\beta},\boldsymbol{t}-[z^{-1}]_{\beta})}
{\tau(\boldsymbol{s},\boldsymbol{t})}z^{s_{\beta}+\delta_{\alpha\beta}-1}e^{\xi(\boldsymbol{t}_{\beta},z)},
\\\Psi_{\alpha\beta}^*(\boldsymbol{s},\boldsymbol{t},z)=&\epsilon_{\alpha\beta}(\boldsymbol{s})\frac{\tau(\boldsymbol{s}-\boldsymbol{e}_{\alpha}+\boldsymbol{e}_{\beta},\boldsymbol{t}+[z^{-1}]_{\beta})}
{\tau(\boldsymbol{s},\boldsymbol{t})}z^{-s_{\beta}+\delta_{\alpha\beta}-1}e^{-\xi(\boldsymbol{t}_{\beta},z)}.
\end{split}
\end{equation}The bilinear identity \eqref{eq12_24_1} is then equivalent to
\begin{equation}\label{eq12_28_6}
\begin{split}
\oint dz \Psi(\boldsymbol{s},\boldsymbol{t},z)\Psi^*(\boldsymbol{s}',\boldsymbol{t}',z)^{T}=0.
\end{split}
\end{equation}
\section{ Differential and Difference Fay Identities}
In this section,     four sets of  differential Fay identities satisfied by the tau function of a multicomponent KP hierarchy are derived from the bilinear identity. This has in fact been considered in \cite{1}. The main difference is that we do not specify a particular component which was called the $0^{\text{th}}$-component in \cite{1}. Two sets of difference Fay identities are also derived from the bilinear relation. It is shown that  the difference Fay identities can in fact be deduced from the differential Fay identities. Therefore they do not contain any new information other than those encoded in the differential Fay identities.

\subsection{Differential Fay Identities}

 The differential Fay identities of a multicomponent KP hierarchy can be divided into four sets:\\
\textbf{DFI}:\; For any $\alpha$,
\begin{equation}\label{eq12_24_5}
\begin{split}
 \frac{\pa_{t_{\alpha 1}}\log\tau\left(\boldsymbol{s}
,\boldsymbol{t}-[\mu^{-1}]_{\alpha}\right) -\pa_{t_{\alpha 1}}\log\tau\left(\boldsymbol{s}
,\boldsymbol{t}-[\nu^{-1}]_{\alpha}\right)}{ \mu-\nu}= \frac{\tau\left(\boldsymbol{s}
,\boldsymbol{t} \right)\tau\left(\boldsymbol{s}, \boldsymbol{t}-[\mu^{-1}]_{\alpha}-[\nu^{-1}]_{\alpha}  \right)}{  \tau\left(\boldsymbol{s}
,\boldsymbol{t} -[\mu^{-1}]_{\alpha}\right)\tau\left(\boldsymbol{s}, \boldsymbol{t}-[\nu^{-1}]_{\alpha}  \right)}-1.
\end{split}
\end{equation}
\textbf{DFII}:\;For any distinct $\alpha$ and $\beta$,
\begin{equation}\label{eq12_24_6}
\begin{split}
&\frac{\pa_{t_{\beta 1}}\log\tau\left(\boldsymbol{s}
,\boldsymbol{t}-[\mu^{-1}]_{\alpha}\right)
  -\pa_{t_{\beta 1}}\log\tau\left(\boldsymbol{s}
,\boldsymbol{t}-[\nu^{-1}]_{\alpha}\right)}{\mu^{-1}-\nu^{-1}}\\=&-\frac{
\tau\left(\boldsymbol{s}
+\boldsymbol{e}_{\alpha}-\boldsymbol{e}_{\beta},\boldsymbol{t} \right)
\tau\left(\boldsymbol{s}-\boldsymbol{e}_{\alpha}+\boldsymbol{e}_{\beta}, \boldsymbol{t}-[\mu^{-1}]_{\alpha}-[\nu^{-1}]_{\alpha}  \right)}{\tau\left(\boldsymbol{s}, \boldsymbol{t} -[\mu^{-1}]_{\alpha}  \right)\tau\left(\boldsymbol{s}, \boldsymbol{t} -[\nu^{-1}]_{\alpha}  \right)}.
\end{split}
\end{equation}
\textbf{DFIII}:\;For any distinct $\alpha$ and $\beta$,
\begin{equation}\label{eq12_24_7}
\begin{split}
&\pa_{t_{\alpha 1}}\log\tau\left(\boldsymbol{s}
+\boldsymbol{e}_{\alpha}-\boldsymbol{e}_{\beta},\boldsymbol{t}-[\nu^{-1}]_{\beta}\right)-\pa_{t_{\alpha 1}}\log\tau\left(\boldsymbol{s}
,\boldsymbol{t}-[\mu^{-1}]_{\alpha}\right)\\=& \mu
 -\mu \frac{\tau\left(\boldsymbol{s}
,\boldsymbol{t}\right)\tau\left(\boldsymbol{s}+\boldsymbol{e}_{\alpha}-\boldsymbol{e}_{\beta}, \boldsymbol{t} -[\mu^{-1}]_{\alpha}-[\nu^{-1}]_{\beta} \right)}{\tau\left(\boldsymbol{s}
,\boldsymbol{t}-[\mu^{-1}]_{\alpha}\right)\tau\left(\boldsymbol{s}+\boldsymbol{e}_{\alpha}-\boldsymbol{e}_{\beta}, \boldsymbol{t} -[\nu^{-1}]_{\beta} \right)}.  \end{split}
\end{equation}
\textbf{DFIV}:\;For any distinct $\alpha,\beta$ and $\kappa$,
\begin{equation}\label{eq12_24_8}
\begin{split}
& \pa_{t_{\kappa 1}}\log\tau\left(\boldsymbol{s},\boldsymbol{t}-[\mu^{-1}]_{\alpha}\right)
-
\pa_{t_{\kappa 1}}\log\tau\left(\boldsymbol{s}
+\boldsymbol{e}_{\alpha}-\boldsymbol{e}_{\beta},\boldsymbol{t}-[\nu^{-1}]_{\beta}\right)
\\
=&-\frac{\epsilon_{\alpha\kappa}(\boldsymbol{s})\epsilon_{\beta\kappa}(\boldsymbol{s})}{\epsilon_{\beta\alpha}(\boldsymbol{s})}\frac{\tau\left(\boldsymbol{s}
+\boldsymbol{e}_{\alpha}-\boldsymbol{e}_{\kappa},\boldsymbol{t} \right)
\tau\left(\boldsymbol{s}-\boldsymbol{e}_{\beta}+\boldsymbol{e}_{\kappa}, \boldsymbol{t}-[\mu^{-1}]_{\alpha}-[\nu^{-1}]_{\beta}  \right)}{\tau\left(\boldsymbol{s}+\boldsymbol{e}_{\alpha}-\boldsymbol{e}_{\beta}, \boldsymbol{t} -[\nu^{-1}]_{\beta}  \right)\tau\left(\boldsymbol{s}, \boldsymbol{t}-[\mu^{-1}]_{\alpha}\right)}.
\end{split}
\end{equation}They are generalizations of the identities (61), (63), (62) and (64) in \cite{1}. For completeness, we give their derivations here.

\begin{proof}For \textbf{DFI},
differentiate  \eqref{eq12_24_2} with respect to $t_{\alpha_1}$ and set
$\boldsymbol{s}'=\boldsymbol{s},$ $\boldsymbol{t}'=\boldsymbol{t}-[\mu^{-1}]_{\alpha}-[\nu^{-1}]_{\alpha}$,
 we have
\begin{equation*}\begin{split}
& \oint dz   \frac{1}{1-\frac{z}{\mu}}\frac{1}{1-\frac{z}{\nu}}\Bigl\{z\tau\left(\boldsymbol{s}
,\boldsymbol{t}-[z^{-1}]_{\alpha}\right)+\pa_{t_{\alpha 1}}\tau\left(\boldsymbol{s}
,\boldsymbol{t}-[z^{-1}]_{\alpha}\right)\Bigr\}
\tau\left(\boldsymbol{s}, \boldsymbol{t}-[\mu^{-1}]_{\alpha}-[\nu^{-1}]_{\alpha} + [z^{-1}]_{\alpha}\right) =0.
\end{split}\end{equation*} Computing the residue, we find that
\begin{equation*}
\begin{split}
&\Bigl\{\mu\tau\left(\boldsymbol{s}
,\boldsymbol{t}-[\mu^{-1}]_{\alpha}\right)+\pa_{t_{\alpha 1}}\tau\left(\boldsymbol{s}
,\boldsymbol{t}-[\mu^{-1}]_{\alpha}\right)\Bigr\}
\tau\left(\boldsymbol{s}, \boldsymbol{t} -[\nu^{-1}]_{\alpha}  \right)-\mu\tau\left(\boldsymbol{s}
,\boldsymbol{t} \right)\tau\left(\boldsymbol{s}, \boldsymbol{t}-[\mu^{-1}]_{\alpha}-[\nu^{-1}]_{\alpha}  \right)\\& -\Bigl\{\nu\tau\left(\boldsymbol{s}
,\boldsymbol{t}-[\nu^{-1}]_{\alpha}\right)+\pa_{t_{\alpha 1}}\tau\left(\boldsymbol{s}
,\boldsymbol{t}-[\nu^{-1}]_{\alpha}\right)\Bigr\}
\tau\left(\boldsymbol{s}, \boldsymbol{t} -[\mu^{-1}]_{\alpha}  \right)
+\nu \tau\left(\boldsymbol{s}
,\boldsymbol{t} \right)\tau\left(\boldsymbol{s}, \boldsymbol{t}-[\mu^{-1}]_{\alpha}-[\nu^{-1}]_{\alpha}  \right)=0,
\end{split}
\end{equation*}which gives \eqref{eq12_24_5} after some rearrangement.
\\
 For \textbf{DFII}, differentiate \eqref{eq12_24_2} with respect to $t_{\beta 1}$ and set $\boldsymbol{s}'=\boldsymbol{s},$ $\boldsymbol{t}'=\boldsymbol{t}-[\mu^{-1}]_{\alpha}-[\nu^{-1}]_{\alpha},$ we have
\begin{equation*}\begin{split}
& \oint dz   \frac{1}{1-\frac{z}{\mu}}\frac{1}{1-\frac{z}{\nu}} \pa_{t_{\beta 1}}\tau\left(\boldsymbol{s}
,\boldsymbol{t}-[z^{-1}]_{\alpha}\right)
\tau\left(\boldsymbol{s}, \boldsymbol{t}-[\mu^{-1}]_{\alpha}-[\nu^{-1}]_{\alpha} + [z^{-1}]_{\alpha}\right)
\\&+\oint dz z^{-2}\left(z \tau\left(\boldsymbol{s}
+\boldsymbol{e}_{\alpha}-\boldsymbol{e}_{\beta},\boldsymbol{t}-[z^{-1}]_{\beta}\right)
+\pa_{t_{\beta 1}}\tau\left(\boldsymbol{s}
+\boldsymbol{e}_{\alpha}-\boldsymbol{e}_{\beta},\boldsymbol{t}-[z^{-1}]_{\beta}\right)\right)
\\&\hspace{2cm}\times\tau\left(\boldsymbol{s}-\boldsymbol{e}_{\alpha}+\boldsymbol{e}_{\beta}, \boldsymbol{t}-[\mu^{-1}]_{\alpha}-[\nu^{-1}]_{\alpha} + [z^{-1}]_{\beta}\right)=0.
\end{split}\end{equation*} This gives
\begin{equation*}\begin{split}
&\frac{\nu\mu}{\nu-\mu} \pa_{t_{\beta 1}}\tau\left(\boldsymbol{s}
,\boldsymbol{t}-[\mu^{-1}]_{\alpha}\right)
\tau\left(\boldsymbol{s}, \boldsymbol{t} -[\nu^{-1}]_{\alpha}  \right) -\frac{\mu\nu}{\nu-\mu}\pa_{t_{\beta 1}}\tau\left(\boldsymbol{s}
,\boldsymbol{t}-[\nu^{-1}]_{\alpha}\right)
\tau\left(\boldsymbol{s}, \boldsymbol{t} -[\mu^{-1}]_{\alpha}  \right)  \\&+\tau\left(\boldsymbol{s}
+\boldsymbol{e}_{\alpha}-\boldsymbol{e}_{\beta},\boldsymbol{t} \right)
\tau\left(\boldsymbol{s}-\boldsymbol{e}_{\alpha}+\boldsymbol{e}_{\beta}, \boldsymbol{t}-[\mu^{-1}]_{\alpha}-[\nu^{-1}]_{\alpha}  \right)=0,
\end{split}
\end{equation*}which is equivalent to \eqref{eq12_24_6}.
\\
For \textbf{DFIII},
differentiate \eqref{eq12_24_1}  with respect to $t_{\alpha_1}$ and set
$\boldsymbol{s}'=\boldsymbol{s},$ $\boldsymbol{t}'=\boldsymbol{t}-[\mu^{-1}]_{\alpha}-[\nu^{-1}]_{\beta},
$ we have
\begin{equation*}\begin{split}
& \epsilon_{\beta\alpha}(\boldsymbol{s})\oint dz  z^{-1} \frac{1}{1-\frac{z}{\mu}} \Bigl\{z\tau\left(\boldsymbol{s}
,\boldsymbol{t}-[z^{-1}]_{\alpha}\right)+\pa_{t_{\alpha 1}}\tau\left(\boldsymbol{s},\boldsymbol{t}-[z^{-1}]_{\alpha}\right)\Bigr\}\\&\hspace{2cm}\times
\tau\left(\boldsymbol{s}+\boldsymbol{e}_{\alpha}-\boldsymbol{e}_{\beta}, \boldsymbol{t}-[\mu^{-1}]_{\alpha}-[\nu^{-1}]_{\beta} + [z^{-1}]_{\alpha}\right)\\
&+\epsilon_{\alpha\beta}(\boldsymbol{s}) \oint dz z^{-1}\frac{1}{1-\frac{z}{\nu}} \pa_{t_{\alpha 1}}\tau\left(\boldsymbol{s}
+\boldsymbol{e}_{\alpha}-\boldsymbol{e}_{\beta},\boldsymbol{t}-[z^{-1}]_{\beta}\right)
\tau\left(\boldsymbol{s}, \boldsymbol{t}-[\mu^{-1}]_{\alpha}-[\nu^{-1}]_{\beta} + [z^{-1}]_{\beta}\right) =0.
\end{split}\end{equation*}Since $$\epsilon_{\beta\alpha}(\boldsymbol{s})=-\epsilon_{\alpha\beta}(\boldsymbol{s}),$$this gives
\begin{equation*}
\begin{split}
& \Bigl\{\mu\tau\left(\boldsymbol{s}
,\boldsymbol{t}-[\mu^{-1}]_{\alpha}\right)+\pa_{t_{\alpha 1}}\tau\left(\boldsymbol{s}
,\boldsymbol{t}-[\mu^{-1}]_{\alpha}\right)\Bigr\}
\tau\left(\boldsymbol{s}+\boldsymbol{e}_{\alpha}-\boldsymbol{e}_{\beta}, \boldsymbol{t} -[\nu^{-1}]_{\beta} \right)\\&-\mu \tau\left(\boldsymbol{s}
,\boldsymbol{t}\right)\tau\left(\boldsymbol{s}+\boldsymbol{e}_{\alpha}-\boldsymbol{e}_{\beta}, \boldsymbol{t} -[\mu^{-1}]_{\alpha}-[\nu^{-1}]_{\beta} \right)
=\pa_{t_{\alpha 1}}\tau\left(\boldsymbol{s}
+\boldsymbol{e}_{\alpha}-\boldsymbol{e}_{\beta},\boldsymbol{t}-[\nu^{-1}]_{\beta}\right)
\tau\left(\boldsymbol{s}, \boldsymbol{t}-[\mu^{-1}]_{\alpha} \right),
\end{split}
\end{equation*}which is equivalent to \eqref{eq12_24_7}.\\
For \text{DFIV}, differentiate \eqref{eq12_24_1} with respect to $t_{\kappa_1}$ and set
$\boldsymbol{s}'=\boldsymbol{s},$ $\boldsymbol{t}'=\boldsymbol{t}-[\mu^{-1}]_{\alpha}-[\nu^{-1}]_{\beta}$,
 we have
\begin{equation*}\begin{split}
&\epsilon_{\beta\alpha}(\boldsymbol{s}) \oint dz  z^{-1} \frac{1}{1-\frac{z}{\mu}}  \pa_{t_{\kappa 1}}\tau\left(\boldsymbol{s},\boldsymbol{t}-[z^{-1}]_{\alpha}\right)
\tau\left(\boldsymbol{s}+\boldsymbol{e}_{\alpha}-\boldsymbol{e}_{\beta}, \boldsymbol{t}-[\mu^{-1}]_{\alpha}-[\nu^{-1}]_{\beta} + [z^{-1}]_{\alpha}\right)\\
&+\epsilon_{\alpha\beta}(\boldsymbol{s}) \oint dz z^{-1}\frac{1}{1-\frac{z}{\nu}} \pa_{t_{\kappa 1}}\tau\left(\boldsymbol{s}
+\boldsymbol{e}_{\alpha}-\boldsymbol{e}_{\beta},\boldsymbol{t}-[z^{-1}]_{\beta}\right)
\tau\left(\boldsymbol{s}, \boldsymbol{t}-[\mu^{-1}]_{\alpha}-[\nu^{-1}]_{\beta} + [z^{-1}]_{\beta}\right) \\&+\epsilon_{\alpha\kappa}(\boldsymbol{s})\epsilon_{\beta\kappa}(\boldsymbol{s})\oint dz z^{-2} \Bigl\{z \tau\left(\boldsymbol{s}
+\boldsymbol{e}_{\alpha}-\boldsymbol{e}_{\kappa},\boldsymbol{t}-[z^{-1}]_{\kappa}\right)+\pa_{t_{\kappa1}}\tau\left(\boldsymbol{s}
+\boldsymbol{e}_{\alpha}-\boldsymbol{e}_{\kappa},\boldsymbol{t}-[z^{-1}]_{\kappa}\right)\Bigr\}\\&\hspace{3cm}\times
\tau\left(\boldsymbol{s}-\boldsymbol{e}_{\beta}+\boldsymbol{e}_{\kappa}, \boldsymbol{t}-[\mu^{-1}]_{\alpha}-[\nu^{-1}]_{\beta} + [z^{-1}]_{\kappa}\right) =0.
\end{split}\end{equation*}This gives
\begin{equation*}
\begin{split}
&\pa_{t_{\kappa 1}}\tau\left(\boldsymbol{s},\boldsymbol{t}-[\mu^{-1}]_{\alpha}\right)
\tau\left(\boldsymbol{s}+\boldsymbol{e}_{\alpha}-\boldsymbol{e}_{\beta}, \boldsymbol{t} -[\nu^{-1}]_{\beta}  \right)-
\pa_{t_{\kappa 1}}\tau\left(\boldsymbol{s}
+\boldsymbol{e}_{\alpha}-\boldsymbol{e}_{\beta},\boldsymbol{t}-[\nu^{-1}]_{\beta}\right)
\tau\left(\boldsymbol{s}, \boldsymbol{t}-[\mu^{-1}]_{\alpha}\right) \\
&+\frac{\epsilon_{\alpha\kappa}(\boldsymbol{s})\epsilon_{\beta\kappa}(\boldsymbol{s})}{\epsilon_{\beta\alpha}(\boldsymbol{s})}\tau\left(\boldsymbol{s}
+\boldsymbol{e}_{\alpha}-\boldsymbol{e}_{\kappa},\boldsymbol{t} \right)
\tau\left(\boldsymbol{s}-\boldsymbol{e}_{\beta}+\boldsymbol{e}_{\kappa}, \boldsymbol{t}-[\mu^{-1}]_{\alpha}-[\nu^{-1}]_{\beta}  \right) =0,
\end{split}
\end{equation*}
 which is equivalent to \eqref{eq12_24_8}.
\end{proof}

\subsection{Difference Fay Identities}
The   difference Fay identities of a multicomponent KP hierarchy can be divided into two sets:\\
\textbf{CFI}:\; For any distinct $\alpha, \beta$, $\lambda$ and $\kappa$,
\begin{equation}\label{eq12_24_4}
\begin{split}
& \epsilon_{\beta\alpha}(\boldsymbol{s}+\boldsymbol{e}_{\lambda} -\boldsymbol{e}_{\kappa})  \tau\left(\boldsymbol{s}
,\boldsymbol{t}\right)
\tau\left(\boldsymbol{s}+\boldsymbol{e}_{\lambda} -\boldsymbol{e}_{\kappa}-\boldsymbol{e}_{\beta}+\boldsymbol{e}_{\alpha}, \boldsymbol{t}-[\mu^{-1}]_{\kappa}\right)\\&  +\epsilon_{\alpha\beta}(\boldsymbol{s}) \tau\left(\boldsymbol{s}+\boldsymbol{e}_{\alpha}-\boldsymbol{e}_{\beta}
,\boldsymbol{t}\right)
\tau\left(\boldsymbol{s}+\boldsymbol{e}_{\lambda} -\boldsymbol{e}_{\kappa}, \boldsymbol{t}-[\mu^{-1}]_{\kappa}\right)\\&+
 \epsilon_{\alpha\kappa}(\boldsymbol{s})\epsilon_{ \beta\kappa}(\boldsymbol{s}+\boldsymbol{e}_{\lambda} -\boldsymbol{e}_{\kappa})\tau\left(\boldsymbol{s}
+\boldsymbol{e}_{\alpha}-\boldsymbol{e}_{\kappa},\boldsymbol{t}-[\mu^{-1}]_{\kappa}\right)
\tau\left(\boldsymbol{s}+\boldsymbol{e}_{\lambda} -\boldsymbol{e}_{\beta}, \boldsymbol{t}\right)=0.
\end{split}
\end{equation}
 \textbf{CFII}:\; For distinct $\alpha,\beta$ and $\lambda$,
\begin{equation}\label{eq12_24_3}
\begin{split}
&\epsilon_{\beta\alpha}(\boldsymbol{s} ) \tau\left(\boldsymbol{s}
,\boldsymbol{t} \right)
\tau\left(\boldsymbol{s}-\boldsymbol{e}_{\beta}+\boldsymbol{e}_{\alpha}, \boldsymbol{t} -[\mu^{-1}]_{\lambda} \right)+\epsilon_{\alpha\beta}(\boldsymbol{s}) \tau\left(\boldsymbol{s}+\boldsymbol{e}_{\alpha}-\boldsymbol{e}_{\beta}
,\boldsymbol{t} \right)
\tau\left(\boldsymbol{s}, \boldsymbol{t}-[\mu^{-1}]_{\lambda} \right)\\
&+\epsilon_{\alpha\lambda}(\boldsymbol{s})\epsilon_{ \beta\lambda}(\boldsymbol{s}) \mu^{  -1} \tau\left(\boldsymbol{s}
+\boldsymbol{e}_{\alpha}-\boldsymbol{e}_{\lambda},\boldsymbol{t}-[\mu^{-1}]_{\lambda}\right)
\tau\left(\boldsymbol{s}-\boldsymbol{e}_{\beta}+\boldsymbol{e}_{\lambda}, \boldsymbol{t} \right)=0.
\end{split}
\end{equation}
\begin{proof}

For \textbf{CFI},
set $\boldsymbol{s}'=\boldsymbol{s}+\boldsymbol{e}_{\lambda} -\boldsymbol{e}_{\kappa}$ and $\boldsymbol{t}'=\boldsymbol{t}-[\mu^{-1}]_{\kappa}$ in the bilinear identity \eqref{eq12_24_1}.  We have
\begin{equation*}
\begin{split}
  & \epsilon_{\beta\alpha}(\boldsymbol{s}+\boldsymbol{e}_{\lambda} -\boldsymbol{e}_{\kappa})\oint dz z^{-1} \tau\left(\boldsymbol{s},\boldsymbol{t}
 -[z^{-1}]_{\alpha}\right)
\tau\left(\boldsymbol{s}+\boldsymbol{e}_{\lambda} -\boldsymbol{e}_{\kappa}-\boldsymbol{e}_{\beta}+\boldsymbol{e}_{\alpha}, \boldsymbol{t}-[\mu^{-1}]_{\kappa}+ [z^{-1}]_{\alpha}\right)\\
&+\epsilon_{\alpha\beta}(\boldsymbol{s}) \oint dz z^{-1} \tau\left(\boldsymbol{s}
+\boldsymbol{e}_{\alpha}-\boldsymbol{e}_{\beta},\boldsymbol{t}-[z^{-1}]_{\beta}\right)
\tau\left(\boldsymbol{s}+\boldsymbol{e}_{\lambda} -\boldsymbol{e}_{\kappa}, \boldsymbol{t}-[\mu^{-1}]_{\kappa}+ [z^{-1}]_{\beta}\right)\\
&+\epsilon_{\alpha\kappa}(\boldsymbol{s})\epsilon_{\beta\kappa}(\boldsymbol{s}+\boldsymbol{e}_{\lambda} -\boldsymbol{e}_{\kappa})\oint dz z^{ -1}\frac{1}{1-\frac{z}{\mu}}\tau\left(\boldsymbol{s}
+\boldsymbol{e}_{\alpha}-\boldsymbol{e}_{\kappa},\boldsymbol{t}-[z^{-1}]_{\kappa}\right)
\tau\left(\boldsymbol{s}+\boldsymbol{e}_{\lambda}  -\boldsymbol{e}_{\beta}, \boldsymbol{t}-[\mu^{-1}]_{\kappa}+ [z^{-1}]_{\kappa}\right)\\
&=0,
\end{split}
\end{equation*}which gives \eqref{eq12_24_4}. \textbf{CFII} is proved in the same way but by taking $\lambda=\kappa$.
\end{proof}

\subsection{Difference Fay Identities From Differential Fay Identities}In this section, we show that the Difference Fay identities \textbf{CFI} and \textbf{CFII} can in fact be derived from the differential Fay identities \textbf{DFII, DFIII} and \textbf{DFIV}.
\begin{proof}
Taking $\nu\rightarrow \infty$ in \eqref{eq12_24_6}, \eqref{eq12_24_7} and \eqref{eq12_24_8}, we find that
\begin{equation}\label{eq2_4_5}
\begin{split}
&\mu\left( \pa_{t_{\beta 1}}\log\tau\left(\boldsymbol{s}
,\boldsymbol{t}-[\mu^{-1}]_{\alpha}\right)
  -\pa_{t_{\beta 1}}\log\tau\left(\boldsymbol{s}
,\boldsymbol{t} \right)\right)=-\frac{
\tau\left(\boldsymbol{s}
+\boldsymbol{e}_{\alpha}-\boldsymbol{e}_{\beta},\boldsymbol{t} \right)
\tau\left(\boldsymbol{s}-\boldsymbol{e}_{\alpha}+\boldsymbol{e}_{\beta}, \boldsymbol{t}-[\mu^{-1}]_{\alpha}   \right)}{\tau\left(\boldsymbol{s}, \boldsymbol{t} -[\mu^{-1}]_{\alpha}  \right)\tau\left(\boldsymbol{s}, \boldsymbol{t}   \right)},
\end{split}
\end{equation}
\begin{equation}\label{eq2_4_6}
\begin{split}
&\pa_{t_{\alpha 1}}\log\tau\left(\boldsymbol{s}
+\boldsymbol{e}_{\alpha}-\boldsymbol{e}_{\beta},\boldsymbol{t}\right)-\pa_{t_{\alpha 1}}\log\tau\left(\boldsymbol{s}
,\boldsymbol{t}-[\mu^{-1}]_{\alpha}\right)= \mu
 -\mu \frac{\tau\left(\boldsymbol{s}
,\boldsymbol{t}\right)\tau\left(\boldsymbol{s}+\boldsymbol{e}_{\alpha}-\boldsymbol{e}_{\beta}, \boldsymbol{t} -[\mu^{-1}]_{\alpha} \right)}{\tau\left(\boldsymbol{s}
,\boldsymbol{t}-[\mu^{-1}]_{\alpha}\right)\tau\left(\boldsymbol{s}+\boldsymbol{e}_{\alpha}-\boldsymbol{e}_{\beta}, \boldsymbol{t}  \right)},  \end{split}
\end{equation}
\begin{equation}\label{eq2_4_7}
\begin{split}
& \pa_{t_{\kappa 1}}\log\tau\left(\boldsymbol{s},\boldsymbol{t}-[\mu^{-1}]_{\alpha}\right)-
\pa_{t_{\kappa 1}}\log\tau\left(\boldsymbol{s}
+\boldsymbol{e}_{\alpha}-\boldsymbol{e}_{\beta},\boldsymbol{t} \right)
\\=& -\frac{\epsilon_{\alpha\kappa}(\boldsymbol{s})\epsilon_{\beta\kappa}(\boldsymbol{s})}{\epsilon_{\beta\alpha}(\boldsymbol{s})}\frac{\tau\left(\boldsymbol{s}
+\boldsymbol{e}_{\alpha}-\boldsymbol{e}_{\kappa},\boldsymbol{t} \right)
\tau\left(\boldsymbol{s}-\boldsymbol{e}_{\beta}+\boldsymbol{e}_{\kappa}, \boldsymbol{t}-[\mu^{-1}]_{\alpha}   \right)}{\tau\left(\boldsymbol{s}+\boldsymbol{e}_{\alpha}-\boldsymbol{e}_{\beta}, \boldsymbol{t}    \right)\tau\left(\boldsymbol{s}, \boldsymbol{t}-[\mu^{-1}]_{\alpha}\right)}.
\end{split}
\end{equation}Taking $\mu\rightarrow \infty$ in \eqref{eq12_24_8} and \eqref{eq2_4_7}, we have
\begin{equation}\label{eq2_4_8}
\begin{split}
& \pa_{t_{\kappa 1}}\log\tau\left(\boldsymbol{s},\boldsymbol{t} \right)
-
\pa_{t_{\kappa 1}}\log\tau\left(\boldsymbol{s}
+\boldsymbol{e}_{\alpha}-\boldsymbol{e}_{\beta},\boldsymbol{t}-[\nu^{-1}]_{\beta}\right)
\\=&-\frac{\epsilon_{\alpha\kappa}(\boldsymbol{s})\epsilon_{\beta\kappa}(\boldsymbol{s})}{\epsilon_{\beta\alpha}(\boldsymbol{s})}\frac{\tau\left(\boldsymbol{s}
+\boldsymbol{e}_{\alpha}-\boldsymbol{e}_{\kappa},\boldsymbol{t} \right)
\tau\left(\boldsymbol{s}-\boldsymbol{e}_{\beta}+\boldsymbol{e}_{\kappa}, \boldsymbol{t} -[\nu^{-1}]_{\beta}  \right)}{\tau\left(\boldsymbol{s}+\boldsymbol{e}_{\alpha}-\boldsymbol{e}_{\beta}, \boldsymbol{t} -[\nu^{-1}]_{\beta}  \right)\tau\left(\boldsymbol{s}, \boldsymbol{t} \right)},
\end{split}
\end{equation}
\begin{equation}\label{eq2_4_9}
\begin{split}
& \pa_{t_{\kappa 1}}\log\tau\left(\boldsymbol{s},\boldsymbol{t} \right)
-\pa_{t_{\kappa 1}}\log\tau\left(\boldsymbol{s}
+\boldsymbol{e}_{\alpha}-\boldsymbol{e}_{\beta},\boldsymbol{t} \right)
= -\frac{\epsilon_{\alpha\kappa}(\boldsymbol{s})\epsilon_{\beta\kappa}(\boldsymbol{s})}{\epsilon_{\beta\alpha}(\boldsymbol{s})}\frac{\tau\left(\boldsymbol{s}
+\boldsymbol{e}_{\alpha}-\boldsymbol{e}_{\kappa},\boldsymbol{t} \right)
\tau\left(\boldsymbol{s}-\boldsymbol{e}_{\beta}+\boldsymbol{e}_{\kappa}, \boldsymbol{t}  \right)}{\tau\left(\boldsymbol{s}+\boldsymbol{e}_{\alpha}-\boldsymbol{e}_{\beta}, \boldsymbol{t}    \right)\tau\left(\boldsymbol{s}, \boldsymbol{t} \right)}.
\end{split}
\end{equation}
Now we can prove \textbf{CFII} \eqref{eq12_24_3}. First, the left hand side of \eqref{eq12_24_3} can be rewritten as
\begin{align*}
&\epsilon_{ \beta\alpha}(\boldsymbol{s}) \tau\left(\boldsymbol{s}+\boldsymbol{e}_{\alpha}-\boldsymbol{e}_{\beta}
,\boldsymbol{t} \right)
\tau\left(\boldsymbol{s}, \boldsymbol{t}-[\mu^{-1}]_{\lambda} \right)\Biggl(\frac{\tau\left(\boldsymbol{s}
,\boldsymbol{t} \right)
\tau\left(\boldsymbol{s}-\boldsymbol{e}_{\beta}+\boldsymbol{e}_{\alpha}, \boldsymbol{t} -[\mu^{-1}]_{\lambda} \right)}{\tau\left(\boldsymbol{s}+\boldsymbol{e}_{\alpha}-\boldsymbol{e}_{\beta}
,\boldsymbol{t} \right)
\tau\left(\boldsymbol{s}, \boldsymbol{t}-[\mu^{-1}]_{\lambda} \right)}-1\\&+\frac{\epsilon_{\alpha\lambda}(\boldsymbol{s})\epsilon_{ \beta\lambda}(\boldsymbol{s})}{\epsilon_{ \beta\alpha}(\boldsymbol{s})} \mu^{  -1} \frac{\tau\left(\boldsymbol{s}
+\boldsymbol{e}_{\alpha}-\boldsymbol{e}_{\lambda},\boldsymbol{t}-[\mu^{-1}]_{\lambda}\right)
\tau\left(\boldsymbol{s}-\boldsymbol{e}_{\beta}+\boldsymbol{e}_{\lambda}, \boldsymbol{t} \right)}{\tau\left(\boldsymbol{s}+\boldsymbol{e}_{\alpha}-\boldsymbol{e}_{\beta}
,\boldsymbol{t} \right)
\tau\left(\boldsymbol{s}, \boldsymbol{t}-[\mu^{-1}]_{\lambda} \right)}\Biggr).
\end{align*}From \eqref{eq2_4_7} with $\alpha\rightarrow \lambda, \kappa\rightarrow \alpha$ and \eqref{eq2_4_5} with $\alpha\rightarrow \lambda, \beta\rightarrow \alpha$, we have
\begin{align*}
&\frac{\tau\left(\boldsymbol{s}
,\boldsymbol{t} \right)
\tau\left(\boldsymbol{s}-\boldsymbol{e}_{\beta}+\boldsymbol{e}_{\alpha}, \boldsymbol{t} -[\mu^{-1}]_{\lambda} \right)}{\tau\left(\boldsymbol{s}+\boldsymbol{e}_{\alpha}-\boldsymbol{e}_{\beta}
,\boldsymbol{t} \right)
\tau\left(\boldsymbol{s}, \boldsymbol{t}-[\mu^{-1}]_{\lambda} \right)}-1 \\&+\frac{\epsilon_{\alpha\lambda}(\boldsymbol{s})\epsilon_{ \beta\lambda}(\boldsymbol{s})}{\epsilon_{ \beta\alpha}(\boldsymbol{s})} \mu^{  -1} \frac{\tau\left(\boldsymbol{s}
+\boldsymbol{e}_{\alpha}-\boldsymbol{e}_{\lambda},\boldsymbol{t}-[\mu^{-1}]_{\lambda}\right)
\tau\left(\boldsymbol{s}-\boldsymbol{e}_{\beta}+\boldsymbol{e}_{\lambda}, \boldsymbol{t} \right)}{\tau\left(\boldsymbol{s}+\boldsymbol{e}_{\alpha}-\boldsymbol{e}_{\beta}
,\boldsymbol{t} \right)
\tau\left(\boldsymbol{s}, \boldsymbol{t}-[\mu^{-1}]_{\lambda} \right)}\\
=&-\frac{\epsilon_{\lambda\alpha}(\boldsymbol{s})\epsilon_{\beta\alpha}(\boldsymbol{s})}{\epsilon_{\beta\lambda}(\boldsymbol{s})}
\frac{\tau\left(\boldsymbol{s}
,\boldsymbol{t} \right)\tau\left(\boldsymbol{s}+\boldsymbol{e}_{\lambda}-\boldsymbol{e}_{\beta}, \boldsymbol{t}    \right) }{\tau\left(\boldsymbol{s}+\boldsymbol{e}_{\alpha}-\boldsymbol{e}_{\beta}
,\boldsymbol{t} \right)\tau\left(\boldsymbol{s}
+\boldsymbol{e}_{\lambda}-\boldsymbol{e}_{\alpha},\boldsymbol{t} \right)}\\&\times\left(\pa_{t_{\alpha 1}}\log\tau\left(\boldsymbol{s},\boldsymbol{t}-[\mu^{-1}]_{\lambda}\right)
-\pa_{t_{\alpha 1}}\log\tau\left(\boldsymbol{s}
+\boldsymbol{e}_{\lambda}-\boldsymbol{e}_{\beta},\boldsymbol{t} \right)\right)-1\\&-\frac{\epsilon_{\alpha\lambda}(\boldsymbol{s})\epsilon_{ \beta\lambda}(\boldsymbol{s})}{\epsilon_{ \beta\alpha}(\boldsymbol{s})}\frac{\tau\left(\boldsymbol{s}
,\boldsymbol{t}\right)
\tau\left(\boldsymbol{s}-\boldsymbol{e}_{\beta}+\boldsymbol{e}_{\lambda}, \boldsymbol{t} \right)}{\tau\left(\boldsymbol{s}+\boldsymbol{e}_{\alpha}-\boldsymbol{e}_{\beta}
,\boldsymbol{t} \right)
\tau\left(\boldsymbol{s}+\boldsymbol{e}_{\lambda}-\boldsymbol{e}_{\alpha}, \boldsymbol{t} \right)}\left( \pa_{t_{\alpha 1}}\log\tau\left(\boldsymbol{s}
,\boldsymbol{t}-[\mu^{-1}]_{\lambda}\right)
  -\pa_{t_{\alpha 1}}\log\tau\left(\boldsymbol{s}
,\boldsymbol{t} \right)\right).
\end{align*}Using \eqref{eq2_4_9} with $\alpha\rightarrow \lambda, \kappa\rightarrow \alpha$, we find that this is equal to
\begin{align*}
&\frac{\pa_{t_{\alpha 1}}\log\tau\left(\boldsymbol{s},\boldsymbol{t}-[\mu^{-1}]_{\lambda}\right)
-\pa_{t_{\alpha 1}}\log\tau\left(\boldsymbol{s}
+\boldsymbol{e}_{\lambda}-\boldsymbol{e}_{\beta},\boldsymbol{t} \right)}{\pa_{t_{\alpha 1}}\log\tau\left(\boldsymbol{s},\boldsymbol{t} \right)
-\pa_{t_{\alpha 1}}\log\tau\left(\boldsymbol{s}
+\boldsymbol{e}_{\lambda}-\boldsymbol{e}_{\beta},\boldsymbol{t} \right)}-1\\&-\frac{\pa_{t_{\alpha 1}}\log\tau\left(\boldsymbol{s}
,\boldsymbol{t}-[\mu^{-1}]_{\lambda}\right)
  -\pa_{t_{\alpha 1}}\log\tau\left(\boldsymbol{s}
,\boldsymbol{t} \right)}{\pa_{t_{\alpha 1}}\log\tau\left(\boldsymbol{s},\boldsymbol{t} \right)
-\pa_{t_{\alpha 1}}\log\tau\left(\boldsymbol{s}
+\boldsymbol{e}_{\lambda}-\boldsymbol{e}_{\beta},\boldsymbol{t} \right)}=0.
\end{align*}This proves \textbf{CFII} \eqref{eq12_24_3}.
For \textbf{CFI} \eqref{eq12_24_4},  the left hand side can be rewritten as
\begin{equation}\label{eq10_26_1}
\begin{split}
&\epsilon_{\beta\alpha}(\boldsymbol{s} )\tau\left(\boldsymbol{s}+\boldsymbol{e}_{\alpha}-\boldsymbol{e}_{\beta}
,\boldsymbol{t}\right)
\tau\left(\boldsymbol{s}+\boldsymbol{e}_{\lambda} -\boldsymbol{e}_{\kappa}, \boldsymbol{t}-[\mu^{-1}]_{\kappa}\right)\\&\times\Biggl\{\frac{\epsilon_{\beta\alpha}(\boldsymbol{s}+\boldsymbol{e}_{\lambda} -\boldsymbol{e}_{\kappa})}{\epsilon_{\beta\alpha}(\boldsymbol{s} )}\frac{ \tau\left(\boldsymbol{s}
,\boldsymbol{t}\right)
\tau\left(\boldsymbol{s}+\boldsymbol{e}_{\lambda} -\boldsymbol{e}_{\kappa}-\boldsymbol{e}_{\beta}+\boldsymbol{e}_{\alpha}, \boldsymbol{t}-[\mu^{-1}]_{\kappa}\right)}{\tau\left(\boldsymbol{s}+\boldsymbol{e}_{\alpha}-\boldsymbol{e}_{\beta}
,\boldsymbol{t}\right)
\tau\left(\boldsymbol{s}+\boldsymbol{e}_{\lambda} -\boldsymbol{e}_{\kappa}, \boldsymbol{t}-[\mu^{-1}]_{\kappa}\right)}-1     \\&+
 \frac{\epsilon_{\alpha\kappa}(\boldsymbol{s})\epsilon_{ \beta\kappa}(\boldsymbol{s}+\boldsymbol{e}_{\lambda} -\boldsymbol{e}_{\kappa})}{\epsilon_{\beta\alpha}(\boldsymbol{s} )}\frac{\tau\left(\boldsymbol{s}
+\boldsymbol{e}_{\alpha}-\boldsymbol{e}_{\kappa},\boldsymbol{t}-[\mu^{-1}]_{\kappa}\right)
\tau\left(\boldsymbol{s}+\boldsymbol{e}_{\lambda} -\boldsymbol{e}_{\beta}, \boldsymbol{t}\right)}{\tau\left(\boldsymbol{s}+\boldsymbol{e}_{\alpha}-\boldsymbol{e}_{\beta}
,\boldsymbol{t}\right)
\tau\left(\boldsymbol{s}+\boldsymbol{e}_{\lambda} -\boldsymbol{e}_{\kappa}, \boldsymbol{t}-[\mu^{-1}]_{\kappa}\right)}\Biggr\}.
\end{split}
\end{equation} Using \eqref{eq2_4_7} with $\alpha\rightarrow \kappa, \kappa\rightarrow \alpha, \boldsymbol{s}\rightarrow \boldsymbol{s}+\boldsymbol{e}_{\lambda}-\boldsymbol{e}_{\kappa}$, we find that
\begin{equation}\label{eq10_26_2}\begin{split}
&\frac{\epsilon_{\beta\alpha}(\boldsymbol{s}+\boldsymbol{e}_{\lambda} -\boldsymbol{e}_{\kappa})}{\epsilon_{\beta\alpha}(\boldsymbol{s} )}\frac{ \tau\left(\boldsymbol{s}
,\boldsymbol{t}\right)
\tau\left(\boldsymbol{s}+\boldsymbol{e}_{\lambda} -\boldsymbol{e}_{\kappa}-\boldsymbol{e}_{\beta}+\boldsymbol{e}_{\alpha}, \boldsymbol{t}-[\mu^{-1}]_{\kappa}\right)}{\tau\left(\boldsymbol{s}+\boldsymbol{e}_{\alpha}-\boldsymbol{e}_{\beta}
,\boldsymbol{t}\right)
\tau\left(\boldsymbol{s}+\boldsymbol{e}_{\lambda} -\boldsymbol{e}_{\kappa}, \boldsymbol{t}-[\mu^{-1}]_{\kappa}\right)}\\
=&-\frac{\epsilon_{\kappa \alpha}(\boldsymbol{s}+\boldsymbol{e}_{\lambda} -\boldsymbol{e}_{\kappa})\epsilon_{ \beta\kappa}(\boldsymbol{s}+\boldsymbol{e}_{\lambda} -\boldsymbol{e}_{\kappa})}{\epsilon_{\beta\alpha}(\boldsymbol{s} )}\frac{ \tau\left(\boldsymbol{s}
,\boldsymbol{t}\right)
\tau\left(\boldsymbol{s}+\boldsymbol{e}_{\lambda} - \boldsymbol{e}_{\beta}, \boldsymbol{t}\right)}{\tau\left(\boldsymbol{s}+\boldsymbol{e}_{\alpha}-\boldsymbol{e}_{\beta}
,\boldsymbol{t}\right)
\tau\left(\boldsymbol{s}+\boldsymbol{e}_{\lambda} -\boldsymbol{e}_{\alpha}, \boldsymbol{t} \right)}\\&\times
\left(\pa_{t_{\alpha 1}}\log\tau\left(\boldsymbol{s}+\boldsymbol{e}_{\lambda}-\boldsymbol{e}_{\kappa},\boldsymbol{t}-[\mu^{-1}]_{\kappa}\right)-
\pa_{t_{\alpha 1}}\log\tau\left(\boldsymbol{s}
+\boldsymbol{e}_{\lambda}-\boldsymbol{e}_{\beta},\boldsymbol{t} \right)\right).
\end{split}\end{equation}Applying \eqref{eq2_4_9} with $\alpha\rightarrow \lambda, \kappa\rightarrow \alpha$, and using the fact that
$$\frac{\epsilon_{\kappa \alpha}(\boldsymbol{s}+\boldsymbol{e}_{\lambda} -\boldsymbol{e}_{\kappa})\epsilon_{ \beta\kappa}(\boldsymbol{s}+\boldsymbol{e}_{\lambda} -\boldsymbol{e}_{\kappa})}{\epsilon_{\beta\lambda}(\boldsymbol{s} )\epsilon_{ \lambda\alpha}(\boldsymbol{s} )}=1,$$
we find that \eqref{eq10_26_2} is equal to
\begin{align*} \frac{\epsilon_{\kappa \alpha}(\boldsymbol{s}+\boldsymbol{e}_{\lambda} -\boldsymbol{e}_{\kappa})\epsilon_{ \beta\kappa}(\boldsymbol{s}+\boldsymbol{e}_{\lambda} -\boldsymbol{e}_{\kappa})}{\epsilon_{\beta\lambda}(\boldsymbol{s} )\epsilon_{ \lambda\alpha}(\boldsymbol{s} )}\frac{\pa_{t_{\alpha 1}}\log\tau\left(\boldsymbol{s}+\boldsymbol{e}_{\lambda}-\boldsymbol{e}_{\kappa},\boldsymbol{t}-[\mu^{-1}]_{\kappa}\right)-
\pa_{t_{\alpha 1}}\log\tau\left(\boldsymbol{s}
+\boldsymbol{e}_{\lambda}-\boldsymbol{e}_{\beta},\boldsymbol{t} \right)}{\pa_{t_{\alpha 1}}\log\tau\left(\boldsymbol{s},\boldsymbol{t} \right)
-\pa_{t_{\alpha 1}}\log\tau\left(\boldsymbol{s}
+\boldsymbol{e}_{\lambda}-\boldsymbol{e}_{\beta},\boldsymbol{t} \right)}.
\end{align*}
 On the other hand, using \eqref{eq2_4_7} with $\alpha\rightarrow \kappa, \kappa\rightarrow \alpha, \beta\rightarrow\lambda, \boldsymbol{s}\rightarrow \boldsymbol{s}+\boldsymbol{e}_{\lambda}-\boldsymbol{e}_{\kappa}$, and the fact that
 \begin{equation*}
 \frac{\epsilon_{\alpha\kappa}(\bs{s})\epsilon_{\lambda\alpha}(\bs{s}+\bs{e}_{\lambda}-\bs{e}_{\kappa})}{\epsilon_{\lambda\kappa}(\bs{s}+\bs{e}_{\lambda}-\bs{e}_{\kappa})}=-1,
 \end{equation*}we find that
\begin{align*}
&\frac{\epsilon_{\alpha\kappa}(\boldsymbol{s})\epsilon_{ \beta\kappa}(\boldsymbol{s}+\boldsymbol{e}_{\lambda} -\boldsymbol{e}_{\kappa})}{\epsilon_{\beta\alpha}(\boldsymbol{s} )}\frac{\tau\left(\boldsymbol{s}
+\boldsymbol{e}_{\alpha}-\boldsymbol{e}_{\kappa},\boldsymbol{t}-[\mu^{-1}]_{\kappa}\right)
\tau\left(\boldsymbol{s}+\boldsymbol{e}_{\lambda} -\boldsymbol{e}_{\beta}, \boldsymbol{t}\right)}{\tau\left(\boldsymbol{s}+\boldsymbol{e}_{\alpha}-\boldsymbol{e}_{\beta}
,\boldsymbol{t}\right)
\tau\left(\boldsymbol{s}+\boldsymbol{e}_{\lambda} -\boldsymbol{e}_{\kappa}, \boldsymbol{t}-[\mu^{-1}]_{\kappa}\right)} \\
=&\frac{\epsilon_{\kappa \alpha}(\boldsymbol{s}+\boldsymbol{e}_{\lambda} -\boldsymbol{e}_{\kappa})\epsilon_{ \beta\kappa}(\boldsymbol{s}+\boldsymbol{e}_{\lambda} -\boldsymbol{e}_{\kappa})}{\epsilon_{\beta\alpha}(\boldsymbol{s} )}\frac{ \tau\left(\boldsymbol{s}
,\boldsymbol{t}\right)
\tau\left(\boldsymbol{s}+\boldsymbol{e}_{\lambda} - \boldsymbol{e}_{\beta}, \boldsymbol{t}\right)}{\tau\left(\boldsymbol{s}+\boldsymbol{e}_{\alpha}-\boldsymbol{e}_{\beta}
,\boldsymbol{t}\right)
\tau\left(\boldsymbol{s}+\boldsymbol{e}_{\lambda} -\boldsymbol{e}_{\alpha}, \boldsymbol{t} \right)}
\\&\times\left(\pa_{t_{\alpha 1}}\log\tau\left(\boldsymbol{s}+\boldsymbol{e}_{\lambda}-\boldsymbol{e}_{\kappa},\boldsymbol{t}-[\mu^{-1}]_{\kappa}\right)-
\pa_{t_{\alpha 1}}\log\tau\left(\boldsymbol{s},\boldsymbol{t} \right)\right)\\
=&-\frac{\pa_{t_{\alpha 1}}\log\tau\left(\boldsymbol{s}+\boldsymbol{e}_{\lambda}-\boldsymbol{e}_{\kappa},\boldsymbol{t}-[\mu^{-1}]_{\kappa}\right)-
\pa_{t_{\alpha 1}}\log\tau\left(\boldsymbol{s},\boldsymbol{t} \right)}{\pa_{t_{\alpha 1}}\log\tau\left(\boldsymbol{s},\boldsymbol{t} \right)
-\pa_{t_{\alpha 1}}\log\tau\left(\boldsymbol{s}
+\boldsymbol{e}_{\lambda}-\boldsymbol{e}_{\beta},\boldsymbol{t} \right)}.
\end{align*}Therefore, \eqref{eq10_26_1} is equal to
\begin{align*}
 &\frac{\pa_{t_{\alpha 1}}\log\tau\left(\boldsymbol{s}+\boldsymbol{e}_{\lambda}-\boldsymbol{e}_{\kappa},\boldsymbol{t}-[\mu^{-1}]_{\kappa}\right)-
\pa_{t_{\alpha 1}}\log\tau\left(\boldsymbol{s}
+\boldsymbol{e}_{\lambda}-\boldsymbol{e}_{\beta},\boldsymbol{t} \right)}{\pa_{t_{\alpha 1}}\log\tau\left(\boldsymbol{s},\boldsymbol{t} \right)
-\pa_{t_{\alpha 1}}\log\tau\left(\boldsymbol{s}
+\boldsymbol{e}_{\lambda}-\boldsymbol{e}_{\beta},\boldsymbol{t} \right)}-1\\&-\frac{\pa_{t_{\alpha 1}}\log\tau\left(\boldsymbol{s}+\boldsymbol{e}_{\lambda}-\boldsymbol{e}_{\kappa},\boldsymbol{t}-[\mu^{-1}]_{\kappa}\right)-
\pa_{t_{\alpha 1}}\log\tau\left(\boldsymbol{s},\boldsymbol{t} \right)}{\pa_{t_{\alpha 1}}\log\tau\left(\boldsymbol{s},\boldsymbol{t} \right)
-\pa_{t_{\alpha 1}}\log\tau\left(\boldsymbol{s}
+\boldsymbol{e}_{\lambda}-\boldsymbol{e}_{\beta},\boldsymbol{t} \right)}=0.
\end{align*}This proves \textbf{CFI} \eqref{eq12_24_4}.\end{proof}

\section{Auxiliary Linear Equations and Lax Equations}

In this section, we derive the auxiliary linear equations for the wave function $\Psi(\boldsymbol{s},\boldsymbol{t},z)$ \eqref{eq12_24_9}. Define the differential operator $$D_{\alpha}(z)=\sum_{j=1}^{\infty}\frac{z^{-j}}{j}\frac{\pa}{\pa t_{\alpha j}}.$$First we have
\begin{proposition}\label{p1} For $1\leq \alpha\leq N$,
the wave function $\Psi(\boldsymbol{s},\boldsymbol{t},z)$ satisfies the following linear equations:
\begin{equation}\label{eq12_24_11}
\begin{split}
\Bigl(1-e^{-D_{\alpha}(\lambda)}\Bigr)\Psi(\boldsymbol{s}
,\boldsymbol{t},z)=\mathfrak{D}(\boldsymbol{s},\boldsymbol{t},\pa,\lambda)\Psi(\boldsymbol{s}
,\boldsymbol{t},z)=\Bigl(\mathfrak{B}(\boldsymbol{s},\boldsymbol{t},\pa,\lambda)E_{\alpha}+E_{\alpha}\mathfrak{C}(\boldsymbol{s},\boldsymbol{t},\lambda)\Bigr) \Psi(\boldsymbol{s}
,\boldsymbol{t},z), 
\end{split}
\end{equation}where
\begin{equation*}
\begin{split}
\mathfrak{B}_{\beta\kappa}(\boldsymbol{s},\boldsymbol{t},\pa,\lambda)=&\left\{\begin{aligned}
-\lambda^{-1}\left(\pa_{t_{\kappa 1}}\log\Psi_{\kappa\kappa}(\boldsymbol{s}
,\boldsymbol{t},\lambda)-\lambda-\pa\right),\hspace{2cm}\text{if}\;\;\beta=\kappa\\
-\epsilon_{\beta\kappa}(\boldsymbol{s}) \frac{\tau\left( \boldsymbol{s}, \boldsymbol{t}\right)
}{\tau\left(\boldsymbol{s}
+\boldsymbol{e}_{\kappa}-\boldsymbol{e}_{\beta},\boldsymbol{t} \right)}\pa_{t_{\beta 1}}\log\Psi_{\kappa\kappa}(\boldsymbol{s}
,\boldsymbol{t},\lambda), \hspace{1cm}\text{if}\;\;\beta\neq \kappa
\end{aligned}\right.;\\
\mathfrak{C}_{\beta\kappa}(\boldsymbol{s},\boldsymbol{t},\lambda)=&\left\{\begin{aligned}
0,\hspace{4cm}\text{if}\;\;\beta=\kappa\\
-\lambda^{-1}\epsilon_{\beta\kappa}(\boldsymbol{s}) \frac{\tau\left(\boldsymbol{s}
+\boldsymbol{e}_{\beta}-\boldsymbol{e}_{\kappa},\boldsymbol{t} \right)}{\tau\left( \boldsymbol{s}, \boldsymbol{t}\right)
}, \hspace{1cm}\text{if}\;\;\beta\neq \kappa
\end{aligned}\right..
\end{split}
\end{equation*}Here $E_{\alpha}$ is the $N\times N$ matrix with $(\alpha,\alpha)$-element equal to one and zero elsewhere,  and $\pa$ is the operator$$\pa=\sum_{\kappa=1}^{N}
\frac{\pa}{\pa t_{\kappa 1}}.$$
\end{proposition}
\begin{proof}
Writing out the components of the equation \eqref{eq12_24_11}, we need to show that
\begin{equation}\label{eq12_24_12}\begin{split}
\Bigl(1-e^{-D_{\alpha}(\lambda)}\Bigr)\Psi_{\alpha\alpha}(\boldsymbol{s}
,\boldsymbol{t},z)=&-\lambda^{-1}\left(\pa_{t_{\alpha 1}}\log\Psi_{\alpha\alpha}(\boldsymbol{s}
,\boldsymbol{t},\lambda)-\lambda-\pa\right)\Psi_{\alpha\alpha}(\boldsymbol{s}
,\boldsymbol{t},z)\\&-\lambda^{-1}\sum_{\substack{1\leq \beta\leq N\\\beta\neq \alpha}}\epsilon_{\alpha\beta}(\boldsymbol{s}) \frac{\tau\left(\boldsymbol{s}
+\boldsymbol{e}_{\alpha}-\boldsymbol{e}_{\beta},\boldsymbol{t} \right)}{\tau\left( \boldsymbol{s}, \boldsymbol{t}\right)
}\Psi_{\beta\alpha}(\boldsymbol{s}
,\boldsymbol{t},z),\end{split}
\end{equation}and if $\beta,\kappa\neq \alpha$,
\begin{equation}\label{eq12_24_13}\begin{split} \Bigl(1-e^{-D_{\alpha}(\lambda)}\Bigr) \Psi_{\beta\alpha}(\boldsymbol{s},\boldsymbol{t},z)=&-\epsilon_{\beta\alpha}(\boldsymbol{s}) \frac{\tau\left( \boldsymbol{s}, \boldsymbol{t}\right)
}{\tau\left(\boldsymbol{s}
+\boldsymbol{e}_{\alpha}-\boldsymbol{e}_{\beta},\boldsymbol{t} \right)}\pa_{t_{\beta 1}}\log\Psi_{\alpha\alpha}(\boldsymbol{s}
,\boldsymbol{t},\lambda) \Psi_{\alpha\alpha}(\boldsymbol{s}
,\boldsymbol{t},z),\end{split}\end{equation}\begin{equation}\label{eq12_24_14}\begin{split}
\Bigl(1-e^{-D_{\alpha}(\lambda)}\Bigr)\Psi_{\alpha\beta}(\boldsymbol{s},\boldsymbol{t},z)= &-\lambda^{-1}\left(\pa_{t_{\alpha 1}}\log\Psi_{\alpha\alpha}(\boldsymbol{s}
,\boldsymbol{t},\lambda)-\lambda-\pa \right)\Psi_{\alpha\beta}(\boldsymbol{s}
,\boldsymbol{t},z)
\\&-\lambda^{-1}\sum_{\substack{1\leq \kappa\leq N\\\kappa\neq \alpha}}\epsilon_{\alpha\kappa}(\boldsymbol{s}) \frac{\tau\left(\boldsymbol{s}
+\boldsymbol{e}_{\alpha}-\boldsymbol{e}_{\kappa},\boldsymbol{t} \right)}{\tau\left( \boldsymbol{s}, \boldsymbol{t}\right)
}\Psi_{\kappa\beta}(\boldsymbol{s}
,\boldsymbol{t},z),\end{split}\end{equation}\begin{equation}\label{eq12_24_15}
\Bigl(1-e^{-D_{\alpha}(\lambda)}\Bigr) \Psi_{\beta\kappa}(\boldsymbol{s},\boldsymbol{t},z)=-\epsilon_{\beta\alpha}(\boldsymbol{s}) \frac{\tau\left( \boldsymbol{s}, \boldsymbol{t}\right)
}{\tau\left(\boldsymbol{s}
+\boldsymbol{e}_{\alpha}-\boldsymbol{e}_{\beta},\boldsymbol{t} \right)}\pa_{t_{\beta 1}}\log\Psi_{\alpha\alpha}(\boldsymbol{s}
,\boldsymbol{t},\lambda) \Psi_{\alpha\kappa}(\boldsymbol{s}
,\boldsymbol{t},z).
\end{equation}
First, we notice that for any $\alpha,\beta,\kappa$, if $\alpha=\kappa$,
\begin{equation*}
\begin{split}
& e^{-D_{\alpha}(\lambda)} \Psi_{\beta\alpha}(\boldsymbol{s}
,\boldsymbol{t},z)= e^{-D_{\alpha}(\lambda)} \left\{\epsilon_{\beta\alpha}(\boldsymbol{s})\frac{\tau(\boldsymbol{s}+
\boldsymbol{e}_{\beta}-\boldsymbol{e}_{\alpha},\boldsymbol{t}-[z^{-1}]_{\alpha})}{\tau(\boldsymbol{s},
\boldsymbol{t})}z^{s_{\alpha}+\delta_{\beta\alpha}-1}e^{\xi(\boldsymbol{t}_{\alpha},z)}\right\}\\
=& \epsilon_{\beta\alpha}(\boldsymbol{s})\frac{\tau(\boldsymbol{s}+\boldsymbol{e}_{\beta}-\boldsymbol{e}_{\alpha},
\boldsymbol{t}-[\lambda^{-1}]_{\alpha}-[z^{-1}]_{\alpha})}
{\tau(\boldsymbol{s},
\boldsymbol{t}-[\lambda^{-1}]_{\alpha})}z^{s_{\alpha}+\delta_{\beta\alpha}-1}\left(1-\frac{z}{\lambda}\right)e^{\xi(\boldsymbol{t}_{\alpha},z)},
\end{split}
\end{equation*}and if $\alpha\neq \kappa$,
\begin{equation*}
\begin{split}
& e^{-D_{\alpha}(\lambda)} \Psi_{\beta\kappa}(\boldsymbol{s}
,\boldsymbol{t},z)= e^{-D_{\alpha}(\lambda)} \left\{\epsilon_{\beta\kappa}(\boldsymbol{s})\frac{\tau(\boldsymbol{s}+
\boldsymbol{e}_{\beta}-\boldsymbol{e}_{\kappa},\boldsymbol{t}-[z^{-1}]_{\kappa})}{\tau(\boldsymbol{s},
\boldsymbol{t})}z^{s_{\kappa}+\delta_{\beta\kappa}-1}e^{\xi(\boldsymbol{t}_{\kappa},z)}\right\}\\
=& \epsilon_{\beta\kappa}(\boldsymbol{s})\frac{\tau(\boldsymbol{s}+
\boldsymbol{e}_{\beta}-\boldsymbol{e}_{\kappa},\boldsymbol{t}-[\lambda^{-1}]_{\alpha}-[z^{-1}]_{\kappa})}
{\tau(\boldsymbol{s},
\boldsymbol{t}-[\lambda^{-1}]_{\alpha})}z^{s_{\kappa}+\delta_{\beta\kappa}-1} e^{\xi(\boldsymbol{t}_{\kappa},z)}.
\end{split}
\end{equation*}On the other hand, for any $\alpha,\beta,\kappa$, we have
\begin{equation*}
\begin{split}
\pa_{t_{\kappa 1}}\log\Psi_{\alpha\beta}(\boldsymbol{s},\boldsymbol{t},z)=\pa_{t_{\kappa 1}}\log\tau(\boldsymbol{s}+\boldsymbol{e}_{\alpha}-\boldsymbol{e}_{\beta},
\boldsymbol{t}-[z^{-1}]_{\beta})-\pa_{t_{\kappa 1}}\log\tau(\boldsymbol{s},\boldsymbol{t})+\delta_{\beta\kappa}z.
\end{split}\end{equation*}
Therefore with $\mu=z$, $\nu=\lambda$, \textbf{DFI} \eqref{eq12_24_5} implies  that
\begin{equation*}
\begin{split}
\pa_{t_{\alpha1}}\log\Psi_{\alpha\alpha}(\boldsymbol{s},\boldsymbol{t},z)- \pa_{t_{\alpha 1}}\log\Psi_{\alpha\alpha}(\boldsymbol{s},\boldsymbol{t},\lambda)
=&(z-\lambda)\frac{\tau\left(\boldsymbol{s}
,\boldsymbol{t} \right)\tau\left(\boldsymbol{s}, \boldsymbol{t}-[\lambda^{-1}]_{\alpha}-[z^{-1}]_{\alpha}  \right)}{  \tau\left(\boldsymbol{s}
,\boldsymbol{t} -[\lambda^{-1}]_{\alpha}\right)\tau\left(\boldsymbol{s}, \boldsymbol{t}-[z^{-1}]_{\alpha}  \right)}.\end{split}\end{equation*}Therefore,
\begin{equation}\label{eq12_25_1}\begin{split}
&\Bigl(\pa_{t_{\alpha 1}}\log\Psi_{\alpha\alpha}(\boldsymbol{s},\boldsymbol{t},z)- \pa_{t_{\alpha 1}}\log\Psi_{\alpha\alpha}(\boldsymbol{s},\boldsymbol{t},\lambda)\Bigr)\Psi_{\alpha\alpha}
(\boldsymbol{s}
,\boldsymbol{t},z)\\=&\Bigl(\pa_{t_{\alpha 1}}\log\Psi_{\alpha\alpha}(\boldsymbol{s},\boldsymbol{t},z)- \pa_{t_{\alpha 1}}\log\Psi_{\alpha\alpha}(\boldsymbol{s},\boldsymbol{t},\lambda)\Bigr)\frac{\tau\left(\boldsymbol{s}, \boldsymbol{t}-[z^{-1}]_{\alpha}  \right)}{\tau\left(\boldsymbol{s}
,\boldsymbol{t} \right)}z^{s_{\alpha}}e^{\xi(\boldsymbol{t}_{\alpha},z)}\\=&-\lambda\left(1-\frac{z}{\lambda}\right)
\frac{ \tau\left(\boldsymbol{s}, \boldsymbol{t}-[\lambda^{-1}]_{\alpha}-[z^{-1}]_{\alpha}  \right)}{  \tau\left(\boldsymbol{s}
,\boldsymbol{t} -[\lambda^{-1}]_{\alpha}\right) }z^{s_{\alpha}}e^{\xi(\boldsymbol{t}_{\alpha},z)}\\
=&-\lambda e^{-D_{\alpha}(\lambda)}\Psi_{\alpha\alpha}(\boldsymbol{s},\boldsymbol{t},z).\end{split}
\end{equation}This is in fact eq. (75) in \cite{1}. Now \eqref{eq2_4_5} with $\mu=z$  implies that
\begin{equation}\label{eq12_25_2}
\begin{split}
\pa_{t_{\beta 1}}\Psi_{\alpha\alpha}(\boldsymbol{s},\boldsymbol{t},z)=&\left[\pa_{t_{\beta 1}}\log\Psi_{\alpha\alpha}(\boldsymbol{s},\boldsymbol{t},z)\right]\Psi_{\alpha\alpha}(\boldsymbol{s},\boldsymbol{t},z) \\
=&  \Bigl(\pa_{t_{\beta 1}}\log\tau\left(\boldsymbol{s}
,\boldsymbol{t}-[z^{-1}]_{\alpha}\right)
  -\pa_{t_{\beta 1}}\log\tau\left(\boldsymbol{s}
,\boldsymbol{t} \right)\Bigr)\frac{\tau\left(\boldsymbol{s}, \boldsymbol{t}-[z^{-1}]_{\alpha}  \right)}{\tau\left(\boldsymbol{s}
,\boldsymbol{t} \right)}z^{s_{\alpha}}e^{\xi(\boldsymbol{t}_{\alpha},z)}\\=&\epsilon_{\alpha\beta}(\boldsymbol{s})\frac{\tau\left(\boldsymbol{s}
+\boldsymbol{e}_{\alpha}-\boldsymbol{e}_{\beta},\boldsymbol{t} \right)}{\tau\left(\boldsymbol{s}
,\boldsymbol{t} \right)}\epsilon_{\beta\alpha}(\boldsymbol{s})\frac{
\tau\left(\boldsymbol{s}-\boldsymbol{e}_{\alpha}+\boldsymbol{e}_{\beta}, \boldsymbol{t}-[z^{-1}]_{\alpha} \right)}{\tau\left(\boldsymbol{s}, \boldsymbol{t}    \right)}z^{s_{\alpha}-1}e^{\xi(\boldsymbol{t}_{\alpha},z)}\\
=&\epsilon_{\alpha\beta}(\boldsymbol{s})\frac{\tau\left(\boldsymbol{s}
+\boldsymbol{e}_{\alpha}-\boldsymbol{e}_{\beta},\boldsymbol{t} \right)}{\tau\left(\boldsymbol{s}
,\boldsymbol{t} \right)}\Psi_{\beta\alpha}(\boldsymbol{s},\boldsymbol{t},z).
\end{split}
\end{equation}
 Using the fact that $$\pa\Psi_{\alpha\alpha}(\boldsymbol{s},\boldsymbol{t},z)=\pa_{t_{\alpha 1}}\Psi_{\alpha\alpha}(\boldsymbol{s},\boldsymbol{t},z)+\sum_{\substack{1\leq \beta\leq N\\\beta\neq \alpha}}\pa_{t_{\beta 1}}
  \Psi_{\alpha\alpha}(\boldsymbol{s},\boldsymbol{t},z),$$we find that \eqref{eq12_25_1} and \eqref{eq12_25_2} together give \eqref{eq12_24_12}.

To prove \eqref{eq12_24_13}, notice that with $\mu=z$, $\nu=\lambda$, \textbf{DFII} \eqref{eq12_24_6}
 implies that for $\beta\neq \alpha$,
\begin{equation*}
\begin{split}
&\pa_{t_{\beta 1}}\log\Psi_{\alpha\alpha}(\boldsymbol{s},\boldsymbol{t},z)- \pa_{t_{\beta 1}}\log\Psi_{\alpha\alpha}(\boldsymbol{s},\boldsymbol{t},\lambda)
\\=&-z^{-1}\left(1-\frac{z}{\lambda}\right)\frac{
\tau\left(\boldsymbol{s}
+\boldsymbol{e}_{\alpha}-\boldsymbol{e}_{\beta},\boldsymbol{t} \right)
\tau\left(\boldsymbol{s}-\boldsymbol{e}_{\alpha}+\boldsymbol{e}_{\beta}, \boldsymbol{t}-[\lambda^{-1}]_{\alpha}-[z^{-1}]_{\alpha}  \right)}{\tau\left(\boldsymbol{s}, \boldsymbol{t} -[\lambda^{-1}]_{\alpha}  \right)\tau\left(\boldsymbol{s}, \boldsymbol{t} -[z^{-1}]_{\alpha}  \right)}.\end{split}\end{equation*}
Therefore,
\begin{equation*}
\begin{split}
&\Bigl(\pa_{t_{\beta 1}}\log\Psi_{\alpha\alpha}(\boldsymbol{s},\boldsymbol{t},z)- \pa_{t_{\beta 1}}\log\Psi_{\alpha\alpha}(\boldsymbol{s},\boldsymbol{t},\lambda)\Bigr)\Psi_{\alpha\alpha}
(\boldsymbol{s}
,\boldsymbol{t},z)\\=&\Bigl(\pa_{t_{\beta 1}}\log\Psi_{\alpha\alpha}(\boldsymbol{s},\boldsymbol{t},z)- \pa_{t_{\beta 1}}\log\Psi_{\alpha\alpha}(\boldsymbol{s},\boldsymbol{t},\lambda)\Bigr)\frac{\tau\left(\boldsymbol{s}, \boldsymbol{t}-[z^{-1}]_{\alpha}  \right)}{\tau\left(\boldsymbol{s}
,\boldsymbol{t} \right)}z^{s_{\alpha}}e^{\xi(\boldsymbol{t}_{\alpha},z)}\\=&\epsilon_{\alpha\beta}(\boldsymbol{s})\frac{\tau\left(\boldsymbol{s}
+\boldsymbol{e}_{\alpha}-\boldsymbol{e}_{\beta},\boldsymbol{t} \right)}{\tau\left(\boldsymbol{s}
,\boldsymbol{t} \right)}\epsilon_{\beta\alpha}(\boldsymbol{s})\frac{
\tau\left(\boldsymbol{s}-\boldsymbol{e}_{\alpha}+\boldsymbol{e}_{\beta}, \boldsymbol{t}-[\lambda^{-1}]_{\alpha}-[z^{-1}]_{\alpha}  \right)}{\tau\left(\boldsymbol{s}, \boldsymbol{t} -[\lambda^{-1}]_{\alpha}  \right)}z^{s_{\alpha}-1}\left(1-\frac{z}{\lambda}\right)e^{\xi(\boldsymbol{t}_{\alpha},z)}\\
=&\epsilon_{\alpha\beta}(\boldsymbol{s})\frac{\tau\left(\boldsymbol{s}
+\boldsymbol{e}_{\alpha}-\boldsymbol{e}_{\beta},\boldsymbol{t} \right)}{\tau\left(\boldsymbol{s}
,\boldsymbol{t} \right)}e^{-D_{\alpha}(\lambda)}\Psi_{\beta\alpha}(\boldsymbol{s},\boldsymbol{t},z).
\end{split}
\end{equation*}Together with \eqref{eq12_25_2}, we find that
\begin{equation*}
\begin{split}
&\epsilon_{\alpha\beta}(\boldsymbol{s})\frac{\tau\left(\boldsymbol{s}
+\boldsymbol{e}_{\alpha}-\boldsymbol{e}_{\beta},\boldsymbol{t} \right)}{\tau\left(\boldsymbol{s}
,\boldsymbol{t} \right)}\Psi_{\beta\alpha}(\boldsymbol{s},\boldsymbol{t},z)-\bigl[\pa_{t_{\beta 1}}\log\Psi_{\alpha\alpha}(\boldsymbol{s},\boldsymbol{t},\lambda)\bigr]\Psi_{\alpha\alpha}
(\boldsymbol{s}
,\boldsymbol{t},z)\\=&\epsilon_{\alpha\beta}(\boldsymbol{s})\frac{\tau\left(\boldsymbol{s}
+\boldsymbol{e}_{\alpha}-\boldsymbol{e}_{\beta},\boldsymbol{t} \right)}{\tau\left(\boldsymbol{s}
,\boldsymbol{t} \right)}e^{-D_{\alpha}(\lambda)}\Psi_{\beta\alpha}(\boldsymbol{s},\boldsymbol{t},z),
\end{split}
\end{equation*}which is equivalent to \eqref{eq12_24_13}.

Now to prove \eqref{eq12_24_14},   \textbf{DFIII} \eqref{eq12_24_7} with $\nu=z$ and $\mu=\lambda$   shows  that for $\alpha\neq \beta$,
\begin{equation*}
\begin{split}
&\left(\pa_{t_{\alpha 1}}\log\Psi_{\alpha\beta}(\boldsymbol{s},\boldsymbol{t},z)-\pa_{t_{\alpha 1}}\log\Psi_{\alpha\alpha}(\boldsymbol{s},\boldsymbol{t},\lambda)\right)\frac{\tau\left(\boldsymbol{s}+\boldsymbol{e}_{\alpha}-\boldsymbol{e}_{\beta}, \boldsymbol{t} -[z^{-1}]_{\beta} \right)}{\tau\left(\boldsymbol{s}
,\boldsymbol{t}\right)}z^{s_{\beta}-1}e^{\xi(\boldsymbol{t}_{\beta},z)}\\=&
 -\lambda \frac{\tau\left(\boldsymbol{s}+\boldsymbol{e}_{\alpha}-\boldsymbol{e}_{\beta}, \boldsymbol{t} -[\lambda^{-1}]_{\alpha}-[z^{-1}]_{\beta} \right)}{\tau\left(\boldsymbol{s}
,\boldsymbol{t}-[\lambda^{-1}]_{\alpha}\right)} z^{s_{\beta}-1}e^{\xi(\boldsymbol{t}_{\beta},z)}. \end{split}
\end{equation*}
This gives
\begin{equation}\label{eq12_25_3}
\Bigl(\pa_{t_{\alpha 1}}-\pa_{t_{\alpha 1}}\log\Psi_{\alpha\alpha}(\boldsymbol{s},\boldsymbol{t},\lambda)\Bigr)\Psi_{\alpha\beta}(\boldsymbol{s},\boldsymbol{t},z)=-\lambda
e^{-D_{\alpha}(\lambda)}\Psi_{\alpha\beta}(\boldsymbol{s},\boldsymbol{t},z).
\end{equation}
Interchanging $\alpha$ and $\beta$, replacing $\bs{s}$ by $\bs{s}+\bs{e}_{\alpha}-\bs{e}_{\beta}$, and setting $\mu=z$ in \eqref{eq2_4_6}, we find that
\begin{equation*}
\pa_{t_{\beta 1}}\log\tau(\boldsymbol{s}+\boldsymbol{e}_{\alpha}-\boldsymbol{e}_{\beta},
\boldsymbol{t}-[z^{-1}]_{\beta})-\pa_{t_{\beta 1}}\log\tau(\boldsymbol{s},\boldsymbol{t})+z
=z\frac{\tau(\boldsymbol{s}+\boldsymbol{e}_{\alpha}-\boldsymbol{e}_{\beta},\boldsymbol{t})\tau(\boldsymbol{s},\boldsymbol{t}-[z^{-1}]_{\beta})}{
\tau(\boldsymbol{s}+\boldsymbol{e}_{\alpha}-\boldsymbol{e}_{\beta},\boldsymbol{t}-[z^{-1}]_{\beta})\tau(\boldsymbol{s},\boldsymbol{t})}.
\end{equation*}This gives
\begin{equation}\label{eq12_25_4}\pa_{t_{\beta 1}}\Psi_{\alpha\beta}(\boldsymbol{s},\boldsymbol{t},z)=\epsilon_{\alpha\beta}(\boldsymbol{s})
\frac{\tau(\boldsymbol{s}+\boldsymbol{e}_{\alpha}-\boldsymbol{e}_{\beta},\boldsymbol{t}) }{
 \tau(\boldsymbol{s},\boldsymbol{t})}\Psi_{\beta\beta}(\boldsymbol{s},\boldsymbol{t},z).\end{equation}
Now set $\nu=z$ and let $\mu\rightarrow \infty$ in \textbf{DFIV} \eqref{eq12_24_8}, we find that for $\kappa\neq \alpha,\beta$,
\begin{equation}\label{eq12_26_8}
\begin{split}
&
\pa_{t_{\kappa 1}}\log\tau\left(\boldsymbol{s}
+\boldsymbol{e}_{\alpha}-\boldsymbol{e}_{\beta},\boldsymbol{t}-[z^{-1}]_{\beta}\right)
 -\pa_{t_{\kappa 1}}\log\tau\left(\boldsymbol{s},\boldsymbol{t} \right)
\\
=&\frac{\epsilon_{\alpha\kappa}(\boldsymbol{s})\epsilon_{\beta\kappa}(\boldsymbol{s})}{\epsilon_{\beta\alpha}(\boldsymbol{s})}
\frac{\tau\left(\boldsymbol{s}
+\boldsymbol{e}_{\alpha}-\boldsymbol{e}_{\kappa},\boldsymbol{t} \right)
\tau\left(\boldsymbol{s}-\boldsymbol{e}_{\beta}+\boldsymbol{e}_{\kappa}, \boldsymbol{t} -[z^{-1}]_{\beta}  \right)}{\tau\left(\boldsymbol{s}+\boldsymbol{e}_{\alpha}-\boldsymbol{e}_{\beta}, \boldsymbol{t} -[z^{-1}]_{\beta}  \right)\tau\left(\boldsymbol{s}, \boldsymbol{t} \right)}.
\end{split}
\end{equation}This gives
\begin{equation}\label{eq12_25_5}\pa_{t_{\kappa 1}}\Psi_{\alpha\beta}(\boldsymbol{s},\boldsymbol{t},z)=\epsilon_{\alpha\kappa}(\boldsymbol{s})
\frac{\tau\left(\boldsymbol{s}
+\boldsymbol{e}_{\alpha}-\boldsymbol{e}_{\kappa},\boldsymbol{t} \right)
 }{ \tau\left(\boldsymbol{s}, \boldsymbol{t} \right)}\Psi_{\kappa\beta}(\boldsymbol{s},\boldsymbol{t},z).\end{equation}Combining together \eqref{eq12_25_3}, \eqref{eq12_25_4} and \eqref{eq12_25_5} prove \eqref{eq12_24_14}.

 For \eqref{eq12_24_15}, consider first the case $\beta=\kappa$. Setting $\mu=z$ and $\nu=\lambda$,   interchanging $\alpha$ and $\beta$, and replacing $\boldsymbol{s}$ by $\boldsymbol{s}+\boldsymbol{e}_{\alpha}-\boldsymbol{e}_{\beta}$ in \textbf{DFIII} \eqref{eq12_24_7}, we find that
\begin{equation}\label{eq10_26_5}
\begin{split}
&\pa_{t_{\beta 1}}\log\tau\left(\boldsymbol{s},\boldsymbol{t}-[\lambda^{-1}]_{\alpha}\right)-\pa_{t_{\beta 1}}\log\tau\left(\boldsymbol{s}
+\boldsymbol{e}_{\alpha}-\boldsymbol{e}_{\beta},\boldsymbol{t}-[z^{-1}]_{\beta}\right)\\=& z
 -z \frac{\tau\left(\boldsymbol{s}+\boldsymbol{e}_{\alpha}-\boldsymbol{e}_{\beta}
,\boldsymbol{t}\right)\tau\left(\boldsymbol{s}, \boldsymbol{t} -[z^{-1}]_{\beta}-[\lambda^{-1}]_{\alpha} \right)}{\tau\left(\boldsymbol{s}
+\boldsymbol{e}_{\alpha}-\boldsymbol{e}_{\beta},\boldsymbol{t}-[z^{-1}]_{\beta}\right)\tau\left(\boldsymbol{s}, \boldsymbol{t} -[\lambda^{-1}]_{\alpha} \right)}.  \end{split}
\end{equation} This gives
\begin{equation*}
\Bigl(\pa_{t_{\beta 1}}\log\Psi_{\alpha\alpha}(\boldsymbol{s},\boldsymbol{t},\lambda)-\pa_{t_{\beta 1}}\log\Psi_{\alpha\beta}(\boldsymbol{s},\boldsymbol{t},z)\Bigr)
\Psi_{\alpha\beta}(\boldsymbol{s},\boldsymbol{t},z)=-\epsilon_{\alpha\beta}(\boldsymbol{s})
\frac{\tau(\boldsymbol{s}+\boldsymbol{e}_{\alpha}-\boldsymbol{e}_{\beta},\boldsymbol{t}) }{
 \tau(\boldsymbol{s},\boldsymbol{t})}e^{-D_{\alpha}(\lambda)}\Psi_{\beta\beta}(\boldsymbol{s},\boldsymbol{t},z).
\end{equation*} Together with \eqref{eq12_25_4}, \eqref{eq12_24_15} is proved when $\beta=\kappa$.
 Finally if $\beta\neq \kappa$,  let $\mu=\lambda$, $\nu=z$ and interchange the role of $\beta$ and $\kappa$ in \textbf{DFIV} \eqref{eq12_24_8} give
\begin{equation*}
\begin{split}
& \pa_{t_{\beta 1}}\log\tau\left(\boldsymbol{s},\boldsymbol{t}-[\lambda^{-1}]_{\alpha}\right)
-
\pa_{t_{\beta 1}}\log\tau\left(\boldsymbol{s}
+\boldsymbol{e}_{\alpha}-\boldsymbol{e}_{\kappa},\boldsymbol{t}-[z^{-1}]_{\kappa}\right)
\\
=&-\frac{\epsilon_{\alpha\beta}(\boldsymbol{s})\epsilon_{\kappa\beta}(\boldsymbol{s})}{\epsilon_{\kappa\alpha}(\boldsymbol{s})}
\frac{\tau\left(\boldsymbol{s}
+\boldsymbol{e}_{\alpha}-\boldsymbol{e}_{\beta},\boldsymbol{t} \right)
\tau\left(\boldsymbol{s}-\boldsymbol{e}_{\kappa}+\boldsymbol{e}_{\beta}, \boldsymbol{t}-[\lambda^{-1}]_{\alpha}-[z^{-1}]_{\kappa}  \right)}{\tau\left(\boldsymbol{s}+\boldsymbol{e}_{\alpha}-\boldsymbol{e}_{\kappa}, \boldsymbol{t} -[z^{-1}]_{\kappa}  \right)\tau\left(\boldsymbol{s}, \boldsymbol{t}-[\lambda^{-1}]_{\alpha}\right)}.
\end{split}
\end{equation*} This shows that
\begin{equation}\label{eq12_25_6}
\begin{split}
&\left(\pa_{t_{\beta 1}}\log\Psi_{\alpha\alpha}(\boldsymbol{s},\boldsymbol{t},\lambda)-\pa_{t_{\beta 1}}\log\Psi_{\alpha\kappa}(\boldsymbol{s},\boldsymbol{t},z)\right)
\Psi_{\alpha\kappa}(\boldsymbol{s},\boldsymbol{t},z)\\=&-\epsilon_{\alpha\beta}(\boldsymbol{s})
\frac{\tau\left(\boldsymbol{s}
+\boldsymbol{e}_{\alpha}-\boldsymbol{e}_{\beta},\boldsymbol{t} \right)}{\tau(\boldsymbol{s},\boldsymbol{t})}e^{-D_{\alpha}(\lambda)}\Psi_{\beta\kappa}(\boldsymbol{s},
\boldsymbol{t},z).
\end{split}\end{equation}Interchanging $\beta$ and $\kappa$ in \eqref{eq12_25_5}, we have
\begin{equation}\label{eq12_25_7}\pa_{t_{\beta 1}}\Psi_{\alpha\kappa}(\boldsymbol{s},\boldsymbol{t},z)=\epsilon_{\alpha\beta}(\boldsymbol{s})
\frac{\tau\left(\boldsymbol{s}
+\boldsymbol{e}_{\alpha}-\boldsymbol{e}_{\beta},\boldsymbol{t} \right)
 }{ \tau\left(\boldsymbol{s}, \boldsymbol{t} \right)}\Psi_{\beta\kappa}(\boldsymbol{s},\boldsymbol{t},z).\end{equation}\eqref{eq12_25_6} and \eqref{eq12_25_7} together give \eqref{eq12_24_15} when $\beta\neq \kappa$.
\end{proof}

By the definition of $\Psi_{\alpha\beta}(\boldsymbol{s},\boldsymbol{t},z)$, we see that it can be written as
\begin{equation}\label{eq12_25_9}
\begin{split}
\Psi_{\alpha\beta}(\boldsymbol{s},\boldsymbol{t},z)=&\left(\delta_{\alpha\beta}+\sum_{j=1}^{\infty} (w_j)_{\alpha,\beta}(
\boldsymbol{s},\boldsymbol{t})z^{-j}\right)z^{s_{\beta}}e^{\xi(\boldsymbol{t}_{\beta},z)}\\
=&W_{\alpha,\beta}(\boldsymbol{s},\boldsymbol{t},\partial)e^{\xi(\boldsymbol{t}_{\beta},z)}\\
=&\hat{W}_{\alpha,\beta}(\boldsymbol{s},\boldsymbol{t},\partial)\pa^{s_{\beta}}e^{\xi(\boldsymbol{t}_{\beta},z)},
\end{split}
\end{equation}where
\begin{equation*}
\begin{split}
W_{\alpha,\beta}(\boldsymbol{s},\boldsymbol{t},\partial)=\left(\delta_{\alpha\beta}+\sum_{j=1}^{\infty} (w_j)_{\alpha,\beta}(
\boldsymbol{s},\boldsymbol{t})\pa^{-j}\right)\pa^{s_{\beta}}=\hat{W}_{\alpha,\beta}(\boldsymbol{s},\boldsymbol{t},\partial)\pa^{s_{\beta}}.
\end{split}
\end{equation*}Now since \begin{equation*}\begin{split}\exp\left(\sum_{\alpha=1}^{N}E_{\alpha}\xi(\boldsymbol{t}_{\alpha},z)\right)=&
\prod_{\alpha=1}^N \exp\left(E_{\alpha}\xi(\boldsymbol{t}_{\alpha},z)\right)= \prod_{\alpha=1}^N\left\{
\mathbb{I}+E_{\alpha} \left(\frac{\xi(\boldsymbol{t}_{\alpha},z)}{1!}+\frac{\xi(\boldsymbol{t}_{\alpha},z)^2}{2!}+\ldots\right)\right\}\\
=&\mathbb{I}+\sum_{\alpha=1}^{N}E_{\alpha} \left(\frac{\xi(\boldsymbol{t}_{\alpha},z)}{1!}+\frac{\xi(\boldsymbol{t}_{\alpha},z)^2}{2!}+\ldots\right)\\
=& \sum_{\alpha=1}^NE_{\alpha}e^{\xi(\boldsymbol{t}_{\alpha},z)}.
\end{split}\end{equation*}Therefore, \eqref{eq12_25_9} can be written in the matrix form
\begin{equation}\label{eq12_28_1}\begin{split}\Psi(\boldsymbol{s}, \boldsymbol{t},z)=&W(\boldsymbol{s},\boldsymbol{t}, \pa)\exp\left(\sum_{\alpha=1}^N E_{\alpha}\xi(\boldsymbol{t}_{\alpha},z)\right)\\=&\hat{W}(\boldsymbol{s},\boldsymbol{t}, \pa)\begin{pmatrix} \pa^{s_{1}}& 0 & \ldots & 0\\
0 & \pa^{s_2} & \ldots & 0\\
\vdots&\vdots& & \vdots\\
0 & 0 & \ldots & \pa^{s_N}\end{pmatrix}\exp\left(\sum_{\alpha=1}^N E_{\alpha}\xi(\boldsymbol{t}_{\alpha},z)\right).
\end{split}
\end{equation}
Using Proposition \ref{p1}, we can find the evolution of the wave function $\Psi(\boldsymbol{s},\boldsymbol{t},z)$ with respect to the time variable.
\begin{proposition}\label{p2}[Linear equations in the $\boldsymbol{t}$-sector]
The wave function $\Psi(\boldsymbol{s},\boldsymbol{t},z)$ satisfies the following linear equations:
\begin{equation*}
\begin{split}
\frac{\pa \Psi(\boldsymbol{s},\boldsymbol{t},z)}{\pa t_{\alpha j}}=B_{\alpha j}(\boldsymbol{s},\boldsymbol{t},\pa)\Psi(\boldsymbol{s},\boldsymbol{t},z),\hspace{1cm}
1\leq\alpha\leq N, j\in\mathbb{N},
\end{split}
\end{equation*} where
\begin{equation}\label{eq12_28_3}
B_{\alpha j}(\boldsymbol{s},\boldsymbol{t},\pa)=\Bigl(W(\boldsymbol{s},\boldsymbol{t}, \pa) E_{\alpha}\pa^j W(\boldsymbol{s},\boldsymbol{t}, \pa)^{-1}\Bigr)_+=\Bigl(\hat{W}(\boldsymbol{s},\boldsymbol{t}, \pa) E_{\alpha}\pa^j \hat{W}(\boldsymbol{s},\boldsymbol{t}, \pa)^{-1}\Bigr)_+.
\end{equation}

\end{proposition}
\begin{proof}
Notice that
\begin{equation}\label{eq10_26_6}
\begin{split}
\sum_{k=1}^{\infty} \frac{\left(1-e^{-D_{\alpha}(\lambda)}\right)^k}{k}=-\log\left(1-\left[1-e^{-D_{\alpha}(\lambda)}\right]\right)=
D_{\alpha}(\lambda)=\sum_{j=1}^{\infty}\frac{\lambda^{-j}}{j}\frac{\pa}{\pa t_{\alpha j}}.
\end{split}
\end{equation}By Proposition \ref{p1},
\begin{equation*}
\left(1-e^{-D_{\alpha}(\lambda)}\right)\Psi(\boldsymbol{s},\boldsymbol{t},z)= \mathfrak{D}(\boldsymbol{s},\boldsymbol{t},\pa, \lambda) \Psi(\boldsymbol{s},\boldsymbol{t},z)
=\left\{\left(\lambda^{-1}\pa \right)E_{\alpha}+\mathfrak{D}_{1,0}(\boldsymbol{s},\boldsymbol{t},\lambda)\right\}\Psi(\boldsymbol{s},\boldsymbol{t},z),
\end{equation*}where $\mathfrak{D}_{1,0}(\boldsymbol{s},\boldsymbol{t},\lambda)$ can be expanded as$$\mathfrak{D}_{1,0}(\boldsymbol{s},\boldsymbol{t},\lambda)=\sum_{j=1}^{\infty}\mathfrak{D}_{1,0;j}(\boldsymbol{s},\boldsymbol{t})\lambda^{-j}.$$Applying again the operator $\left(1-e^{-D_{\alpha}(\lambda)}\right)$, we find that
\begin{equation*}
\begin{split}
&\left(1-e^{-D_{\alpha}(\lambda)}\right)^2\Psi(\boldsymbol{s},\boldsymbol{t},z)\\=&\Bigl\{\left(\lambda^{-1}\pa \right)E_{\alpha}+\mathfrak{D}_{1,0}(\boldsymbol{s},\boldsymbol{t}-[\lambda^{-1}]_{\alpha},\lambda)\Bigr\}
\left(1-e^{-D_{\alpha}(\lambda)}\right)\Psi(\boldsymbol{s},\boldsymbol{t},z)
\\&+\Bigl\{\mathfrak{D}_{1,0}(\boldsymbol{s},\boldsymbol{t},\lambda)-\mathfrak{D}_{1,0}(\boldsymbol{s},\boldsymbol{t}-[\lambda^{-1}]_{\alpha},\lambda)\Bigr\}
\Psi(\boldsymbol{s},\boldsymbol{t},z)\\
=&\left\{\Bigl[\left(\lambda^{-1}\pa \right)E_{\alpha}+\mathfrak{D}_{1,0}(\boldsymbol{s},\boldsymbol{t}-[\lambda^{-1}]_{\alpha},\lambda)\Bigr]\Bigl[\left(\lambda^{-1}\pa \right)E_{\alpha}+\mathfrak{D}_{1,0}(\boldsymbol{s},\boldsymbol{t},\lambda)\Bigr]\right.
\\&\left.+\Bigl[\mathfrak{D}_{1,0}(\boldsymbol{s},\boldsymbol{t},\lambda)
-\mathfrak{D}_{1,0}(\boldsymbol{s},\boldsymbol{t}-[\lambda^{-1}]_{\alpha},\lambda)\Bigr]\right\}\Psi(\boldsymbol{s},\boldsymbol{t},z)\\
=&\Bigl\{\left(\lambda^{-1}\pa \right)^2E_{\alpha}+\mathfrak{D}_{2,1}(\boldsymbol{s},\boldsymbol{t},\lambda)\pa+
\mathfrak{D}_{2,0}(\boldsymbol{s},\boldsymbol{t},\lambda)\Bigr\}\Psi(\boldsymbol{s},\boldsymbol{t},z),
\end{split}
\end{equation*}where for $i=0,1$,
\begin{equation*}
\mathfrak{D}_{2,i}(\boldsymbol{s},\boldsymbol{t},\lambda)=\sum_{j=2}^{\infty}\mathfrak{D}_{2,i;j}(\boldsymbol{s},\boldsymbol{t})\lambda^{-j}.
\end{equation*}By induction, one can show that
\begin{equation*}
\left(1-e^{-D_{\alpha}(\lambda)}\right)^k\Psi(\boldsymbol{s},\boldsymbol{t},z)=\left(\sum_{i=0}^k \mathfrak{D}_{k,i}(\boldsymbol{s},\boldsymbol{t},\lambda)\pa^{i}\right)
\Psi(\boldsymbol{s},\boldsymbol{t},z),
\end{equation*}where
\begin{equation*}\begin{split}
\mathfrak{D}_{k,k}(\boldsymbol{s},\boldsymbol{t},\lambda) =&  \lambda^{-k}E_{\alpha}\hspace{0.5cm}\text{and}\hspace{0.5cm}
\mathfrak{D}_{k,i}(\boldsymbol{s},\boldsymbol{t},\lambda)=\sum_{j=k}^{\infty}\mathfrak{D}_{k,i;j}(\boldsymbol{s},\boldsymbol{t})\lambda^{-j}, \hspace{1cm} 0\leq i\leq k-1.
\end{split}\end{equation*}Therefore, \eqref{eq10_26_6} gives
\begin{equation*}
\begin{split}
\sum_{j=1}^{\infty}\frac{\lambda^{-j}}{j}\frac{\pa\Psi(\boldsymbol{s},\boldsymbol{t},z)}{\pa t_{\alpha j}}
=&
\sum_{k=1}^{\infty}\frac{1}{k}\left(\sum_{j=k}^{\infty}\sum_{i=0}^{k}\mathfrak{D}_{k,i;j}(\boldsymbol{s},\boldsymbol{t})
\lambda^{-j}\pa^{i}\right)\Psi(\boldsymbol{s},\boldsymbol{t},z)\\
=&\sum_{j=1}^{\infty} \lambda^{-j}\left(\sum_{k=1}^j\frac{1}{k}\sum_{i=0}^k \mathfrak{D}_{k,i;j}(\boldsymbol{s},\boldsymbol{t})\pa^{i}\right)\Psi(\boldsymbol{s},\boldsymbol{t},z).
\end{split}
\end{equation*}Comparing coefficients of $\lambda^{j}$, we find that
\begin{equation*}
\begin{split}
\frac{\pa\Psi(\boldsymbol{s},\boldsymbol{t},z)}{\pa t_{\alpha j}}=&j\left(\sum_{k=1}^j\frac{1}{k}\sum_{i=0}^k \mathfrak{D}_{k,i;j}(\boldsymbol{s},\boldsymbol{t})\pa^{i}\right)\Psi(\boldsymbol{s},\boldsymbol{t},z)\\
=&j\left(\sum_{k=1}^j\frac{1}{k} \mathfrak{D}_{k,0;j}(\boldsymbol{s},\boldsymbol{t})+\sum_{i=1}^j\left[\sum_{k=i}^j
\frac{\mathfrak{D}_{k,i;j}(\boldsymbol{s},\boldsymbol{t})}{k}\pa^{i}\right]\right)\Psi(\boldsymbol{s},\boldsymbol{t},z).
\end{split}\end{equation*}
Notice that$$B_{\alpha j}(\boldsymbol{s},\boldsymbol{t},\pa)=j\left(\sum_{k=1}^j\frac{1}{k} \mathfrak{D}_{k,0;j}(\boldsymbol{s},\boldsymbol{t})+\sum_{i=1}^j\left[\sum_{k=i}^j
\frac{\mathfrak{D}_{k,i;j}(\boldsymbol{s},\boldsymbol{t})}{k}\pa^{i}\right]\right)$$ is a differential operator in $\pa$ with leading term
$$\mathfrak{D}_{j,j;j}(\boldsymbol{s},\boldsymbol{t})\pa^j=E_{\alpha}\pa^j.$$
On the other hand, \eqref{eq12_28_1} implies that
\begin{equation*}\begin{split}
\frac{\pa \Psi(\boldsymbol{s},\boldsymbol{t},z)}{\pa t_{\alpha j}}=&\left(\frac{\pa \hat{W}(\boldsymbol{s},\boldsymbol{t},\pa)}{\pa t_{\alpha j}}\hat{W}(\boldsymbol{s},\boldsymbol{t},\pa)^{-1}\right)\Psi(\boldsymbol{s},\boldsymbol{t},z)\\&+\hat{W}(\boldsymbol{s},\boldsymbol{t}, \pa)\begin{pmatrix} \pa^{s_{1}}& 0 & \ldots & 0\\
0 & \pa^{s_2} & \ldots & 0\\
\vdots&\vdots& & \vdots\\
0 & 0 & \ldots & \pa^{s_N}\end{pmatrix}E_{\alpha}z^j\exp\left(\sum_{\alpha=1}^N E_{\alpha}\xi(\boldsymbol{t}_{\alpha},z)\right)\\
=&\left(\frac{\pa \hat{W}(\boldsymbol{s},\boldsymbol{t},\pa)}{\pa t_{\alpha j}}\hat{W}(\boldsymbol{s},\boldsymbol{t},\pa)^{-1}\right)\Psi(\boldsymbol{s},\boldsymbol{t},z)+\left(\hat{W}(\boldsymbol{s},\boldsymbol{t}, \pa)E_{\alpha}\pa^j
\hat{W}(\boldsymbol{s},\boldsymbol{t},\pa)^{-1}\right)\Psi(\boldsymbol{s},\boldsymbol{t},z).
\end{split}\end{equation*}Therefore,
\begin{equation}\label{eq12_28_2}
B_{\alpha j}(\boldsymbol{s},\boldsymbol{t},\pa)=\left(\frac{\pa \hat{W}(\boldsymbol{s},\boldsymbol{t},\pa)}{\pa t_{\alpha j}}\hat{W}(\boldsymbol{s},\boldsymbol{t},\pa)^{-1}\right)+\left(\hat{W}(\boldsymbol{s},\boldsymbol{t}, \pa)E_{\alpha}\pa^j
\hat{W}(\boldsymbol{s},\boldsymbol{t},\pa)^{-1}\right).
\end{equation}Since $$\left(\frac{\pa \hat{W}(\boldsymbol{s},\boldsymbol{t},\pa)}{\pa t_{\alpha j}}\hat{W}(\boldsymbol{s},\boldsymbol{t},\pa)^{-1}\right)$$ is a pseudodifferential operator in $\pa$ that only contains negative powers of $\pa$, but $B_{\alpha j}(\boldsymbol{s},\boldsymbol{t},\pa)$ is a differential operator, comparing both sides of \eqref{eq12_28_2} proves \eqref{eq12_28_3}.

As a side remark, we also deduce from \eqref{eq12_28_2} that
\begin{equation}\label{eq12_28_4}
\left(\frac{\pa \hat{W}(\boldsymbol{s},\boldsymbol{t},\pa)}{\pa t_{\alpha j}}\hat{W}(\boldsymbol{s},\boldsymbol{t},\pa)^{-1}\right)=-\left(\hat{W}(\boldsymbol{s},\boldsymbol{t}, \pa)E_{\alpha}\pa^j
\hat{W}(\boldsymbol{s},\boldsymbol{t},\pa)^{-1}\right)_-.
\end{equation}
\end{proof}

Next we consider the evolution of the wave function $\Psi(\bs{s},\bs{t},z)$ with respect to the $\bs{s}$ variable.
\begin{proposition}\label{p3}[Linear equations in the $\boldsymbol{s}$-sector]
The wave function $\Psi(\boldsymbol{s},\boldsymbol{t},z)$ satisfies the following linear equations: For any distinct $\alpha$ and $\beta$,
\begin{equation}\label{eq12_25_10}
\begin{split}
 \Psi(\boldsymbol{s}+\boldsymbol{e}_{\alpha}-\boldsymbol{e}_{\beta},\boldsymbol{t},z) = P_{\alpha,\beta}(\boldsymbol{s},\boldsymbol{t},\pa)\Psi(\boldsymbol{s},\boldsymbol{t},z)=
 \left(E_{\alpha}\pa+ \mathfrak{G}(\boldsymbol{s},\boldsymbol{t})\right)\Psi(\boldsymbol{s},\boldsymbol{t},z),
\end{split}
\end{equation} where
\begin{equation*}\begin{split}
\mathfrak{G}(\boldsymbol{s},\boldsymbol{t})=&\mathfrak{H}(\boldsymbol{s}+\boldsymbol{e}_{\alpha}-\boldsymbol{e}_{\beta},\boldsymbol{t})E_{\alpha}
-E_{\alpha}\mathfrak{H}(\boldsymbol{s},\boldsymbol{t})+\sum_{\substack{1\leq\gamma\leq N\\\gamma\neq\alpha,\beta}}E_{\gamma},\\
\mathfrak{H}_{\lambda\kappa}(\boldsymbol{s},\boldsymbol{t})=&\left\{\begin{aligned}
-\pa_{t_{\kappa 1}}\log\tau(\boldsymbol{s},\boldsymbol{t}), \hspace{1cm} &\text{if}\;\;\lambda=\kappa\\
\epsilon_{\lambda\kappa}(\boldsymbol{s})\frac{\tau(\boldsymbol{s}
+\boldsymbol{e}_{\lambda}-\boldsymbol{e}_{\kappa},\boldsymbol{t})}{\tau(\boldsymbol{s},\boldsymbol{t})}
,\hspace{0.5cm} &\text{if}\;\;\lambda\neq \kappa
\end{aligned}\right. .\end{split}
\end{equation*}

\end{proposition}
\begin{proof}
Writing out the components of \eqref{eq12_25_10}, we need to show that
\begin{equation}\label{eq12_25_11}
\begin{split}
\Psi_{\alpha\gamma}(\boldsymbol{s}+\boldsymbol{e}_{\alpha}-\boldsymbol{e}_{\beta},\boldsymbol{t},z)=&\Bigl(\pa-\pa_{t_{\alpha 1}}\log\tau(\boldsymbol{s}+\boldsymbol{e}_{\alpha}-\boldsymbol{e}_{\beta},\boldsymbol{t})+\pa_{t_{\alpha 1}}\log\tau(\boldsymbol{s},\boldsymbol{t})\Bigr)\Psi_{\alpha\gamma}
(\boldsymbol{s},\boldsymbol{t},z)\\&-\sum_{\substack{1\leq \kappa\leq N\\\kappa\neq \alpha}}
\epsilon_{\alpha\kappa}(\boldsymbol{s})\frac{\tau(\boldsymbol{s}+\boldsymbol{e}_{\alpha}-\boldsymbol{e}_{\kappa}
,\boldsymbol{t})}{\tau(\boldsymbol{s},\boldsymbol{t})}\Psi_{\kappa\gamma}(\boldsymbol{s},\boldsymbol{t},z);
\end{split}
\end{equation}
and if $\lambda\neq \alpha$,
\begin{equation}\label{eq12_25_12}
\begin{split}
\Psi_{\lambda\gamma}(\boldsymbol{s}+\boldsymbol{e}_{\alpha}-\boldsymbol{e}_{\beta},\boldsymbol{t},z)=&
\epsilon_{\lambda\alpha}(\boldsymbol{s}+\boldsymbol{e}_{\alpha}-\boldsymbol{e}_{\beta})\frac{\tau(\boldsymbol{s}
+\boldsymbol{e}_{\lambda}-\boldsymbol{e}_{\beta},\boldsymbol{t})}{\tau(\boldsymbol{s}+\boldsymbol{e}_{\alpha}-\boldsymbol{e}_{\beta},\boldsymbol{t})}
\Psi_{\alpha\gamma}(\boldsymbol{s},\boldsymbol{t},z)+\left(1-\delta_{\lambda\beta}\right)
\Psi_{\lambda\gamma}(\boldsymbol{s},\boldsymbol{t},z).
\end{split}
\end{equation}
If $\gamma=\alpha$, \eqref{eq12_25_11} is equivalent to
\begin{equation}\label{eq12_26_1}
\begin{split}
&z\frac{\tau(\boldsymbol{s}+\boldsymbol{e}_{\alpha}-\boldsymbol{e}_{\beta},\boldsymbol{t}-[z^{-1}]_{\alpha})\tau(\boldsymbol{s},\boldsymbol{t})}
{\tau(\boldsymbol{s},\boldsymbol{t}-[z^{-1}]_{\alpha})\tau(\boldsymbol{s}+\boldsymbol{e}_{\alpha}-\boldsymbol{e}_{\beta},\boldsymbol{t} )}
\\=&\pa_{t_{\alpha 1}} \log\tau(\boldsymbol{s},\boldsymbol{t}-[z^{-1}]_{\alpha}) +z+\sum_{\substack{1\leq \kappa\leq N\\\kappa\neq\alpha}}
\pa_{t_{\kappa 1}}\log\Psi_{\alpha\alpha}(\boldsymbol{s},\boldsymbol{t},z)
\\&-\pa_{t_{\alpha 1}}\log\tau(\boldsymbol{s}+\boldsymbol{e}_{\alpha}-\boldsymbol{e}_{\beta},\boldsymbol{t} )
 -\sum_{\substack{1\leq\kappa\leq N\\\kappa\neq \alpha}}\epsilon_{\alpha\kappa}(\boldsymbol{s}) \frac{\tau(\boldsymbol{s}+\boldsymbol{e}_{\alpha}-\boldsymbol{e}_{\kappa},
\boldsymbol{t} )}{\tau(\boldsymbol{s},\boldsymbol{t})}\frac{\Psi_{\kappa\alpha}(\boldsymbol{s},\boldsymbol{t},z)}{\Psi_{\alpha\alpha}(\boldsymbol{s},\boldsymbol{t},z)}.
\end{split}
\end{equation}Setting $\mu=z$ in \eqref{eq2_4_6}, we have
\begin{equation}\label{eq12_26_2}
\begin{split}
&\pa_{t_{\alpha 1}}\log\tau\left(\boldsymbol{s}
+\boldsymbol{e}_{\alpha}-\boldsymbol{e}_{\beta},\boldsymbol{t} \right)-\pa_{t_{\alpha 1}}\log\tau\left(\boldsymbol{s}
,\boldsymbol{t}-[z^{-1}]_{\alpha}\right)= z
 -z \frac{\tau\left(\boldsymbol{s}
,\boldsymbol{t}\right)\tau\left(\boldsymbol{s}+\boldsymbol{e}_{\alpha}-\boldsymbol{e}_{\beta}, \boldsymbol{t} -[z^{-1}]_{\alpha} \right)}{\tau\left(\boldsymbol{s}
,\boldsymbol{t}-[z^{-1}]_{\alpha}\right)\tau\left(\boldsymbol{s}+\boldsymbol{e}_{\alpha}-\boldsymbol{e}_{\beta}, \boldsymbol{t}  \right)}.  \end{split}
\end{equation}On the other hand, \eqref{eq12_25_2} shows that
\begin{equation}\label{eq12_26_3}
\begin{split}
 \sum_{\substack{1\leq \kappa\leq N\\\kappa\neq\alpha}}\pa_{t_{\kappa 1}}\log \Psi_{\alpha\alpha}(\boldsymbol{s},\boldsymbol{t}, z)
= \sum_{\substack{1\leq\kappa\leq N\\\kappa\neq \alpha}}\epsilon_{\alpha\kappa}(\boldsymbol{s}) \frac{\tau(\boldsymbol{s}+\boldsymbol{e}_{\alpha}-\boldsymbol{e}_{\kappa},
\boldsymbol{t} )}{\tau(\boldsymbol{s},\boldsymbol{t})}\frac{\Psi_{\kappa\alpha}(\boldsymbol{s},\boldsymbol{t},z)}{\Psi_{\alpha\alpha}(\boldsymbol{s},\boldsymbol{t},z)}.
\end{split}
\end{equation}Eqs. \eqref{eq12_26_2} and \eqref{eq12_26_3} together prove \eqref{eq12_26_1}.

If $\gamma=\beta$, \eqref{eq12_25_11} is equivalent to
\begin{equation}\label{eq12_26_4}
\begin{split}
&-z^{-1}\frac{\tau(\boldsymbol{s}+2\boldsymbol{e}_{\alpha}-2\boldsymbol{e}_{\beta},\boldsymbol{t}-[z^{-1}]_{\beta})\tau(\boldsymbol{s},\boldsymbol{t})}
{\tau(\boldsymbol{s}+\boldsymbol{e}_{\alpha}-\boldsymbol{e}_{\beta},\boldsymbol{t}-[z^{-1}]_{\beta})\tau(\boldsymbol{s}+\boldsymbol{e}_{\alpha}-\boldsymbol{e}_{\beta},\boldsymbol{t} )}
\\&=\pa_{t_{\alpha 1}} \log\tau(\boldsymbol{s}+\boldsymbol{e}_{\alpha}-\boldsymbol{e}_{\beta},\boldsymbol{t}-[z^{-1}]_{\beta})   +\sum_{\substack{1\leq \kappa\leq N\\\kappa\neq\alpha}}
\pa_{t_{\kappa 1}}\log\Psi_{\alpha\beta}(\boldsymbol{s},\boldsymbol{t},z)
 -\pa_{t_{\alpha 1}}\log\tau(\boldsymbol{s}+\boldsymbol{e}_{\alpha}-\boldsymbol{e}_{\beta},\boldsymbol{t} )
 \\&-\sum_{\substack{1\leq\kappa\leq N\\\kappa\neq \alpha}}\epsilon_{\alpha\kappa}(\boldsymbol{s}) \frac{\tau(\boldsymbol{s}+\boldsymbol{e}_{\alpha}-\boldsymbol{e}_{\kappa},
\boldsymbol{t} )}{\tau(\boldsymbol{s},\boldsymbol{t})}
\frac{\Psi_{\kappa\beta}(\boldsymbol{s},\boldsymbol{t},z)}{\Psi_{\alpha\beta}(\boldsymbol{s},\boldsymbol{t},z)}.
\end{split}
\end{equation}Interchanging $\alpha$ and $\beta$, replacing $\boldsymbol{s}$ with $\boldsymbol{s}+\boldsymbol{e}_{\alpha}-\boldsymbol{e}_{\beta}$, and setting $\mu=z$ in \eqref{eq2_4_5} give
\begin{equation}\label{eq12_26_5}\begin{split}
&z\Bigl(\pa_{t_{\alpha 1}} \log\tau(\boldsymbol{s}+\boldsymbol{e}_{\alpha}-\boldsymbol{e}_{\beta},\boldsymbol{t}-[z^{-1}]_{\beta}) -\pa_{t_{\alpha 1}}\log\tau(\boldsymbol{s}+\boldsymbol{e}_{\alpha}-\boldsymbol{e}_{\beta},\boldsymbol{t} )\Bigr)\\=&-\frac{\tau(\boldsymbol{s}+2\boldsymbol{e}_{\alpha}-2\boldsymbol{e}_{\beta},\boldsymbol{t}-[z^{-1}]_{\beta})\tau(\boldsymbol{s},\boldsymbol{t})}
{\tau(\boldsymbol{s}+\boldsymbol{e}_{\alpha}-\boldsymbol{e}_{\beta},\boldsymbol{t}-[z^{-1}]_{\beta})\tau(\boldsymbol{s}+\boldsymbol{e}_{\alpha}-\boldsymbol{e}_{\beta},\boldsymbol{t} )}.
\end{split}\end{equation}On the other hand, \eqref{eq12_25_4} and \eqref{eq12_25_5} give
\begin{equation}\label{eq12_26_6}
\sum_{\substack{1\leq \kappa\leq N\\\kappa\neq\alpha}}
\pa_{t_{\kappa 1}}\log\Psi_{\alpha\beta}(\boldsymbol{s},\boldsymbol{t},z)=\sum_{\substack{1\leq\kappa\leq N\\\kappa\neq \alpha}}\epsilon_{\alpha\kappa}(\boldsymbol{s}) \frac{\tau(\boldsymbol{s}+\boldsymbol{e}_{\alpha}-\boldsymbol{e}_{\kappa},
\boldsymbol{t} )}{\tau(\boldsymbol{s},\boldsymbol{t})}
\frac{\Psi_{\kappa\beta}(\boldsymbol{s},\boldsymbol{t},z)}{\Psi_{\alpha\beta}(\boldsymbol{s},\boldsymbol{t},z)}.
\end{equation}Eqs. \eqref{eq12_26_5} and \eqref{eq12_26_6} together prove \eqref{eq12_26_4}.

If $\gamma\neq \alpha,\beta$,
\eqref{eq12_25_11} is equivalent to
\begin{equation}\label{eq12_26_7}
\begin{split}
&\frac{\epsilon_{\alpha\gamma}(\boldsymbol{s}+\boldsymbol{e}_{\alpha}-\boldsymbol{e}_{\beta})}{\epsilon_{\alpha\gamma}(\boldsymbol{s})}
\frac{\tau(\boldsymbol{s}+2\boldsymbol{e}_{\alpha}- \boldsymbol{e}_{\beta}-\boldsymbol{e}_{\gamma},\boldsymbol{t}-[z^{-1}]_{\gamma})\tau(\boldsymbol{s},\boldsymbol{t})}
{\tau(\boldsymbol{s}+\boldsymbol{e}_{\alpha}-\boldsymbol{e}_{\gamma},\boldsymbol{t}-[z^{-1}]_{\gamma})\tau(\boldsymbol{s}+\boldsymbol{e}_{\alpha}-\boldsymbol{e}_{\beta},\boldsymbol{t} )}
\\=&\pa_{t_{\alpha 1}} \log\tau(\boldsymbol{s}+\boldsymbol{e}_{\alpha}-\boldsymbol{e}_{\gamma},\boldsymbol{t}-[z^{-1}]_{\gamma})  +\sum_{\substack{1\leq \kappa\leq N\\\kappa\neq\alpha}}
\pa_{t_{\kappa 1}}\log\Psi_{\alpha\gamma}(\boldsymbol{s},\boldsymbol{t},z)
\\&-\pa_{t_{\alpha 1}}\log\tau(\boldsymbol{s}+\boldsymbol{e}_{\alpha}-\boldsymbol{e}_{\beta},\boldsymbol{t} )
 -\sum_{\substack{1\leq\kappa\leq N\\\kappa\neq \alpha}}\epsilon_{\alpha\kappa}(\boldsymbol{s}) \frac{\tau(\boldsymbol{s}+\boldsymbol{e}_{\alpha}-\boldsymbol{e}_{\kappa},
\boldsymbol{t} )}{\tau(\boldsymbol{s},\boldsymbol{t})}
\frac{\Psi_{\kappa\gamma}(\boldsymbol{s},\boldsymbol{t},z)}{\Psi_{\alpha\gamma}(\boldsymbol{s},\boldsymbol{t},z)}.
\end{split}
\end{equation}Replacing $\alpha$ with $\gamma$,   $\kappa$ with $\alpha$,   $\boldsymbol{s}$ with $\boldsymbol{s}+\boldsymbol{e}_{\alpha}-\boldsymbol{e}_{\gamma}$ and setting $\mu=z$ in \eqref{eq2_4_7} gives
\begin{equation}\label{eq12_26_9}
\begin{split}
&
\pa_{t_{\alpha 1}}\log\tau\left(\boldsymbol{s}
+\boldsymbol{e}_{\alpha}-\boldsymbol{e}_{\gamma},\boldsymbol{t}-[z^{-1}]_{\gamma}\right)
 -\pa_{t_{\alpha 1}}\log\tau\left(\boldsymbol{s}+\boldsymbol{e}_{\alpha}-\boldsymbol{e}_{\beta},\boldsymbol{t} \right)
\\
=&-\frac{\epsilon_{\beta\alpha}(\boldsymbol{s+\boldsymbol{e}_{\alpha}-\boldsymbol{e}_{\gamma}})\epsilon_{\gamma\alpha}(\boldsymbol{s}
+\boldsymbol{e}_{\alpha}-\boldsymbol{e}_{\gamma})}{\epsilon_{\beta\gamma}(\boldsymbol{s}+\boldsymbol{e}_{\alpha}-\boldsymbol{e}_{\gamma})}
\frac{\tau\left(\boldsymbol{s},\boldsymbol{t} \right)
\tau\left(\boldsymbol{s}+2\boldsymbol{e}_{\alpha}- \boldsymbol{e}_{\beta}-\boldsymbol{e}_{\gamma}, \boldsymbol{t} -[z^{-1}]_{\gamma}  \right)}{\tau\left(\boldsymbol{s}+\boldsymbol{e}_{\alpha}-\boldsymbol{e}_{\gamma}, \boldsymbol{t} -[z^{-1}]_{\gamma}  \right)\tau\left(\boldsymbol{s}+
\boldsymbol{e}_{\alpha}-\boldsymbol{e}_{\beta}, \boldsymbol{t} \right)}.
\end{split}
\end{equation}Now one can prove directly that
$$-\frac{\epsilon_{\beta\alpha}(\boldsymbol{s+\boldsymbol{e}_{\alpha}-\boldsymbol{e}_{\gamma}})\epsilon_{\gamma\alpha}(\boldsymbol{s}
+\boldsymbol{e}_{\alpha}-\boldsymbol{e}_{\gamma})}{\epsilon_{\beta\gamma}(\boldsymbol{s}+\boldsymbol{e}_{\alpha}-\boldsymbol{e}_{\gamma})}
=\frac{\epsilon_{\alpha\gamma}(\boldsymbol{s}+\boldsymbol{e}_{\alpha}-\boldsymbol{e}_{\beta})}{\epsilon_{\alpha\gamma}(\boldsymbol{s})}.$$Therefore \eqref{eq12_26_9} together with \eqref{eq12_26_6} (with $\beta$ replaced by $\gamma$) imply \eqref{eq12_26_7}. This completes the proof of \eqref{eq12_25_11}.

If $\lambda=\beta$, \eqref{eq12_25_12} is equivalent to
\begin{equation}\label{eq12_26_10}
\begin{split}
&\epsilon_{\beta\gamma}(\boldsymbol{s}+\boldsymbol{e}_{\alpha}-\boldsymbol{e}_{\beta})\frac{\tau(\boldsymbol{s}+\boldsymbol{e}_{\alpha}-\boldsymbol{e}_{\gamma}
,\boldsymbol{t}-[z^{-1}]_{\gamma})}{\tau(\boldsymbol{s}+\boldsymbol{e}_{\alpha}-\boldsymbol{e}_{\beta},\boldsymbol{t})}\\=&
\epsilon_{\beta\alpha}(\boldsymbol{s}+\boldsymbol{e}_{\alpha}-\boldsymbol{e}_{\beta})\epsilon_{\alpha\gamma}(\boldsymbol{s})\frac{\tau(\boldsymbol{s},\boldsymbol{t})}{\tau(\boldsymbol{s}+\boldsymbol{e}_{\alpha}-\boldsymbol{e}_{\beta},\boldsymbol{t})}
\frac{\tau(\boldsymbol{s}+\boldsymbol{e}_{\alpha}-\boldsymbol{e}_{\gamma}
,\boldsymbol{t}-[z^{-1}]_{\gamma})}{\tau(\boldsymbol{s},\boldsymbol{t})}.
\end{split}
\end{equation}This is a tautology since $$\epsilon_{\beta\gamma}(\boldsymbol{s}+\boldsymbol{e}_{\alpha}-\boldsymbol{e}_{\beta})=\epsilon_{\beta\alpha}(\boldsymbol{s}+\boldsymbol{e}_{\alpha}-\boldsymbol{e}_{\beta})\epsilon_{\alpha\gamma}(\boldsymbol{s}).$$

If $\lambda\neq \alpha$ or $\beta$, $\gamma=\alpha$, \eqref{eq12_25_12} is equivalent to
\begin{equation}\label{eq12_26_11}
\begin{split}
&z \frac{\tau(\boldsymbol{s}+\boldsymbol{e}_{\lambda}-\boldsymbol{e}_{\beta},\boldsymbol{t}-[z^{-1}]_{\alpha})}{\tau(\boldsymbol{s}
+\boldsymbol{e}_{\alpha}-\boldsymbol{e}_{\beta},\boldsymbol{t})}\\=&z
 \frac{\tau(\boldsymbol{s}
+\boldsymbol{e}_{\lambda}-\boldsymbol{e}_{\beta},\boldsymbol{t})}{\tau(\boldsymbol{s}+\boldsymbol{e}_{\alpha}-\boldsymbol{e}_{\beta},\boldsymbol{t})}
\frac{\tau(\boldsymbol{s},t-[z^{-1}]_{\alpha})}{\tau(\boldsymbol{s},\boldsymbol{t})}+
\frac{\epsilon_{\lambda\alpha}(\boldsymbol{s})}{\epsilon_{\lambda\alpha}(\boldsymbol{s}
+\boldsymbol{e}_{\alpha}-\boldsymbol{e}_{\beta})}\frac{\tau(\boldsymbol{s}+\boldsymbol{e}_{\lambda}-\boldsymbol{e}_{\alpha},t-[z^{-1}]_{\alpha})}{\tau(\boldsymbol{s},\boldsymbol{t})}.
\end{split}
\end{equation}To prove this, interchange $\alpha$ and $\lambda$ in \textbf{CFII} \eqref{eq12_24_3} and set $\mu=z$. This gives
\begin{equation} \label{eq12_27_1}
\begin{split}
&\epsilon_{\beta\lambda}(\boldsymbol{s} ) \tau\left(\boldsymbol{s}
,\boldsymbol{t} \right)
\tau\left(\boldsymbol{s}-\boldsymbol{e}_{\beta}+\boldsymbol{e}_{\lambda}, \boldsymbol{t} -[z^{-1}]_{\alpha} \right)+\epsilon_{\lambda\beta}(\boldsymbol{s}) \tau\left(\boldsymbol{s}+\boldsymbol{e}_{\lambda}-\boldsymbol{e}_{\beta}
,\boldsymbol{t} \right)
\tau\left(\boldsymbol{s}, \boldsymbol{t}-[z^{-1}]_{\alpha} \right)\\
&+\epsilon_{\lambda\alpha}(\boldsymbol{s})\epsilon_{ \beta\alpha}(\boldsymbol{s}) z^{  -1} \tau\left(\boldsymbol{s}
+\boldsymbol{e}_{\lambda}-\boldsymbol{e}_{\alpha},\boldsymbol{t}-[z^{-1}]_{\alpha}\right)
\tau\left(\boldsymbol{s}-\boldsymbol{e}_{\beta}+\boldsymbol{e}_{\alpha}, \boldsymbol{t} \right)=0.
\end{split}
\end{equation}Since
$$\frac{ \epsilon_{ \beta\alpha}(\boldsymbol{s})}{\epsilon_{\lambda\beta}(\boldsymbol{s})}=\frac{1}{\epsilon_{\lambda\alpha}(\boldsymbol{s}
+\boldsymbol{e}_{\alpha}-\boldsymbol{e}_{\beta})},$$ one finds that \eqref{eq12_27_1} is equivalent to \eqref{eq12_26_11}.

If $\lambda\neq \alpha$ or $\beta$, $\gamma=\beta$, \eqref{eq12_25_12} is equivalent to
\begin{equation}\label{eq12_27_3}
\begin{split}&\epsilon_{ \lambda\beta}(\boldsymbol{s}+\boldsymbol{e}_{\alpha}-\boldsymbol{e}_{\beta}) z^{-1}\frac{\tau(\boldsymbol{s}+\boldsymbol{e}_{\alpha}+\boldsymbol{e}_{\lambda}-2\boldsymbol{e}_{\beta},\boldsymbol{t}-[z^{-1}]_{\beta})}{\tau(\boldsymbol{s}
+\boldsymbol{e}_{\alpha}-\boldsymbol{e}_{\beta},\boldsymbol{t})}\\=& \epsilon_{ \lambda\alpha}(\boldsymbol{s}+\boldsymbol{e}_{\alpha}-\boldsymbol{e}_{\beta})\epsilon_{\alpha\beta}(\boldsymbol{s})
 \frac{\tau(\boldsymbol{s}
+\boldsymbol{e}_{\lambda}-\boldsymbol{e}_{\beta},\boldsymbol{t})}{\tau(\boldsymbol{s}+\boldsymbol{e}_{\alpha}-\boldsymbol{e}_{\beta},\boldsymbol{t})}
\frac{\tau(\boldsymbol{s}+\boldsymbol{e}_{\alpha}-\boldsymbol{e}_{\beta},t-[z^{-1}]_{\beta})}{\tau(\boldsymbol{s},\boldsymbol{t})}\\&+
 \epsilon_{\lambda\beta}(\boldsymbol{s}) \frac{\tau(\boldsymbol{s}+\boldsymbol{e}_{\lambda}-\boldsymbol{e}_{\beta},t-[z^{-1}]_{\beta})}{\tau(\boldsymbol{s},\boldsymbol{t})}.
\end{split}
\end{equation}
To prove this, interchange $\beta$ and $\lambda$, replacing $\boldsymbol{s}$ with $\boldsymbol{s}+\boldsymbol{e}_{\lambda}-\boldsymbol{e}_{\beta}$ in \textbf{CFII} \eqref{eq12_24_3} and set $\mu=z$. This gives
\begin{equation} \label{eq12_27_2}
\begin{split}
& \tau\left(\boldsymbol{s}+\boldsymbol{e}_{\lambda}-\boldsymbol{e}_{\beta},\boldsymbol{t} \right)
\tau\left(\boldsymbol{s}-\boldsymbol{e}_{\beta}+\boldsymbol{e}_{\alpha}, \boldsymbol{t} -[z^{-1}]_{\beta} \right)- \tau\left(\boldsymbol{s}+\boldsymbol{e}_{\alpha}-\boldsymbol{e}_{\beta}
,\boldsymbol{t} \right)
\tau\left(\boldsymbol{s}+\boldsymbol{e}_{\lambda}-\boldsymbol{e}_{\beta}, \boldsymbol{t}-[z^{-1}]_{\beta} \right)\\
&+\frac{\epsilon_{\alpha\beta}(\boldsymbol{s}+\boldsymbol{e}_{\lambda}-\boldsymbol{e}_{\beta})\epsilon_{\lambda \beta}(\boldsymbol{s}+\boldsymbol{e}_{\lambda}-\boldsymbol{e}_{\beta}) }{\epsilon_{\lambda\alpha}(\boldsymbol{s}+\boldsymbol{e}_{\lambda}-\boldsymbol{e}_{\beta} ) }z^{  -1} \tau\left(\boldsymbol{s}
+\boldsymbol{e}_{\alpha}+\boldsymbol{e}_{\lambda}-2\boldsymbol{e}_{\beta},\boldsymbol{t}-[z^{-1}]_{\beta}\right)
\tau\left(\boldsymbol{s}, \boldsymbol{t} \right)=0.
\end{split}
\end{equation}On can show that
\begin{equation}\label{eq12_27_5}\epsilon_{ \lambda\alpha}(\boldsymbol{s}+\boldsymbol{e}_{\alpha}-\boldsymbol{e}_{\beta})\epsilon_{\alpha\beta}(\boldsymbol{s})=-
 \epsilon_{\lambda\beta}(\boldsymbol{s}),\hspace{1cm}\frac{\epsilon_{ \lambda\beta}(\boldsymbol{s}+\boldsymbol{e}_{\alpha}-\boldsymbol{e}_{\beta})}
 {\epsilon_{\lambda\beta}(\boldsymbol{s})}=\frac{\epsilon_{\alpha\beta}(\boldsymbol{s}+\boldsymbol{e}_{\lambda}-\boldsymbol{e}_{\beta})\epsilon_{\lambda \beta}(\boldsymbol{s}+\boldsymbol{e}_{\lambda}-\boldsymbol{e}_{\beta}) }{\epsilon_{\lambda\alpha}(\boldsymbol{s}+\boldsymbol{e}_{\lambda}-\boldsymbol{e}_{\beta} ) }.\end{equation}Therefore \eqref{eq12_27_2} is equivalent to \eqref{eq12_27_3}.

If $\lambda\neq \alpha$ or $\beta$, $\gamma=\lambda$, \eqref{eq12_25_12} is equivalent to
\begin{equation}\label{eq12_27_4}
\begin{split}&z\frac{\tau(\boldsymbol{s}+\boldsymbol{e}_{\alpha}-\boldsymbol{e}_{\beta},\boldsymbol{t}-[z^{-1}]_{\lambda})}{\tau(\boldsymbol{s}
+\boldsymbol{e}_{\alpha}-\boldsymbol{e}_{\beta},\boldsymbol{t})}\\=& \epsilon_{ \lambda\alpha}(\boldsymbol{s}+\boldsymbol{e}_{\alpha}-\boldsymbol{e}_{\beta})\epsilon_{\alpha\lambda}(\boldsymbol{s})
 \frac{\tau(\boldsymbol{s}
+\boldsymbol{e}_{\lambda}-\boldsymbol{e}_{\beta},\boldsymbol{t})}{\tau(\boldsymbol{s}+\boldsymbol{e}_{\alpha}-\boldsymbol{e}_{\beta},\boldsymbol{t})}
\frac{\tau(\boldsymbol{s}+\boldsymbol{e}_{\alpha}-\boldsymbol{e}_{\lambda},t-[z^{-1}]_{\lambda})}{\tau(\boldsymbol{s},\boldsymbol{t})}+z
 \frac{\tau(\boldsymbol{s},t-[z^{-1}]_{\lambda})}{\tau(\boldsymbol{s},\boldsymbol{t})}.
\end{split}
\end{equation} The first identity in \eqref{eq12_27_5} implies that   \textbf{CFII} \eqref{eq12_24_3} with $\mu=z$ is equivalent to \eqref{eq12_27_4}.

If $\lambda\neq \alpha$ or $\beta$, $\gamma\neq \alpha,\beta$ or $\lambda$, \eqref{eq12_25_12} is equivalent to
\begin{equation}\label{eq12_27_6}
\begin{split}
&\epsilon_{\lambda\gamma}(\boldsymbol{s}+\boldsymbol{e}_{\alpha}-\boldsymbol{e}_{\beta})\frac{\tau(\boldsymbol{s}+\boldsymbol{e}_{\alpha}
+\boldsymbol{e}_{\lambda}-\boldsymbol{e}_{\beta}-\boldsymbol{e}_{\gamma},\boldsymbol{t}-[z^{-1}]_{\gamma})}{\tau(
\boldsymbol{s}+\boldsymbol{e}_{\alpha}-\boldsymbol{e}_{\beta},\boldsymbol{t})}
\\=&\epsilon_{\lambda\alpha}(\boldsymbol{s}+\boldsymbol{e}_{\alpha}-\boldsymbol{e}_{\beta})
\epsilon_{\alpha\gamma}(\boldsymbol{s})\frac{\tau(\boldsymbol{s}
+\boldsymbol{e}_{\lambda}-\boldsymbol{e}_{\beta},\boldsymbol{t})}{\tau(\boldsymbol{s}+\boldsymbol{e}_{\alpha}-\boldsymbol{e}_{\beta},\boldsymbol{t})}
\frac{\tau(\boldsymbol{s}+\boldsymbol{e}_{\alpha}
-\boldsymbol{e}_{\gamma},\boldsymbol{t}-[z^{-1}]_{\gamma})}{\tau(
\boldsymbol{s},\boldsymbol{t})}\\&+\epsilon_{\lambda\gamma}(\boldsymbol{s})
\frac{\tau(\boldsymbol{s}+\boldsymbol{e}_{\lambda}
-\boldsymbol{e}_{\gamma},\boldsymbol{t}-[z^{-1}]_{\gamma})}{\tau(
\boldsymbol{s},\boldsymbol{t})}.\end{split}
\end{equation}One can show directly that
\begin{equation*}
\frac{\epsilon_{\lambda\alpha}(\boldsymbol{s}+\boldsymbol{e}_{\alpha}-\boldsymbol{e}_{\beta})}
{\epsilon_{\lambda\gamma}(\boldsymbol{s}+\boldsymbol{e}_{\alpha}-\boldsymbol{e}_{\beta})}=-
\frac{\epsilon_{\beta\gamma}(\boldsymbol{s}+\boldsymbol{e}_{\lambda}-\boldsymbol{e}_{\gamma})}
{\epsilon_{\beta\alpha}(\boldsymbol{s}+\boldsymbol{e}_{\lambda}-\boldsymbol{e}_{\gamma})},\hspace{1cm}
\frac{\epsilon_{\lambda\gamma}(\boldsymbol{s})}
{\epsilon_{\lambda\gamma}(\boldsymbol{s}+\boldsymbol{e}_{\alpha}-\boldsymbol{e}_{\beta})}=
-\frac{\epsilon_{\alpha\beta}(\boldsymbol{s})}
{\epsilon_{\beta\alpha}(\boldsymbol{s}+\boldsymbol{e}_{\lambda}-\boldsymbol{e}_{\gamma})}.
\end{equation*}Therefore \eqref{eq12_27_6} is implied immediately by \textbf{CFI} \eqref{eq12_24_4} (with $\kappa$ replaced by $\gamma$.). This completes the proof of \eqref{eq12_25_12} and also the proof of the proposition.

\end{proof}

In Proposition \ref{p2} and Proposition \ref{p3}, we have proved the following auxiliary linear equations
\begin{equation}\label{eq12_28_5}
\begin{split}
\frac{\pa \Psi(\boldsymbol{s},\boldsymbol{t},z)}{\pa t_{\alpha j}}=&B_{\alpha j}(\boldsymbol{s},\boldsymbol{t},\pa)\Psi(\boldsymbol{s},\boldsymbol{t},z),\hspace{1cm}
1\leq\alpha\leq N, j\in\mathbb{N},
\\\Psi(\boldsymbol{s}+\boldsymbol{e}_{\alpha}-\boldsymbol{e}_{\beta},\boldsymbol{t},z) = & P_{\alpha,\beta}(\boldsymbol{s},\boldsymbol{t},\pa)\Psi(\boldsymbol{s},\boldsymbol{t},z),\hspace{1cm}1\leq\alpha,\beta\leq N, \alpha\neq\beta,
\end{split}
\end{equation}where $B_{\alpha j}(\boldsymbol{s},\boldsymbol{t},\pa)$ given by \eqref{eq12_28_1} is a differential operator in $\pa$ of order $j$, and
$P_{\alpha,\beta}(\boldsymbol{s},\boldsymbol{t},\pa)$ is a first order differential operator in $\pa$. Define the $N\times N$ matrix operators $L$ and $R_{\alpha}$, $\alpha=1,\ldots,N$, by
\begin{equation*}\begin{split}L(\boldsymbol{s},\boldsymbol{t},\pa)=&\hat{W}(\boldsymbol{s},\boldsymbol{t},\pa)\pa\hat{W}(\boldsymbol{s},\boldsymbol{t},\pa)^{-1}=
W(\boldsymbol{s},\boldsymbol{t},\pa)\pa W(\boldsymbol{s},\boldsymbol{t},\pa)^{-1}
=\pa+\sum_{n=1}^{\infty}u_{n}(\boldsymbol{s},\boldsymbol{t})\pa^{-n},\\
R_{\alpha}(\boldsymbol{s},\boldsymbol{t},\pa)=&\hat{W}(\boldsymbol{s},\boldsymbol{t},\pa)E_{\alpha}\hat{W}(\boldsymbol{s},\boldsymbol{t},\pa)^{-1}
=W(\boldsymbol{s},\boldsymbol{t},\pa)E_{\alpha} W(\boldsymbol{s},\boldsymbol{t},\pa)^{-1}=
E_{\alpha}+\sum_{n=1}^{\infty}u_{\alpha n}(\boldsymbol{s},\boldsymbol{t})\pa^{-n}.\end{split}\end{equation*}
Then it is straightforward to verify that
$$L R_{\alpha} =R_{\alpha} L,
\hspace{1cm}R_{\alpha} R_{\beta} =\delta_{\alpha\beta}R_{\alpha},\hspace{1cm}\sum_{\alpha=1}^N R_{\alpha}=\mathbf{1},$$ $$B_{\alpha j}(\boldsymbol{s},\boldsymbol{t},\pa)= \Bigl(\hat{W}(\boldsymbol{s},\boldsymbol{t}, \pa) E_{\alpha}\pa^j \hat{W}(\boldsymbol{s},\boldsymbol{t}, \pa)^{-1}\Bigr)_+=(L^jR_{\alpha})_+,$$and
\eqref{eq12_28_1},
\eqref{eq12_28_4} and \eqref{eq12_28_5} imply that
\begin{equation*}
\begin{split}
\frac{\pa W(\boldsymbol{s},\boldsymbol{t},\pa)}{\pa t_{\alpha j}}W(\boldsymbol{s},\boldsymbol{t},\pa)^{-1}=&-\left(W(\boldsymbol{s},\boldsymbol{t},\pa)\pa^jE_{\alpha}W(\boldsymbol{s},\boldsymbol{t},\pa)^{-1}\right)_-,\\
W(\boldsymbol{s}+\boldsymbol{e}_{\alpha}-\boldsymbol{e}_{\beta},\boldsymbol{t},\pa)=&P_{\alpha,\beta}(\boldsymbol{s},\boldsymbol{t},\pa)W(\boldsymbol{s},\boldsymbol{t},\pa).
\end{split}
\end{equation*}Therefore,
\begin{equation}\label{eq10_28_3}
\begin{split}
&\frac{\pa L(\boldsymbol{s},\boldsymbol{t},\pa)}{\pa t_{\gamma j}}=[B_{\gamma j}(\boldsymbol{s},\boldsymbol{t},\pa),L(\boldsymbol{s},\boldsymbol{t},\pa)],\\
&\frac{\pa R_{\beta}(\boldsymbol{s},\boldsymbol{t},\pa)}{\pa_{t_{\gamma j}}}=[B_{\gamma j}(\boldsymbol{s},\boldsymbol{t},\pa),R_{\beta}(\boldsymbol{s},\boldsymbol{t},\pa)],\\
&L( \boldsymbol{s} +\boldsymbol{e}_{\alpha}-e_{\beta},\boldsymbol{t},\pa)P_{\alpha,\beta}\left(\boldsymbol{s},\boldsymbol{t},\pa\right)=
P_{\alpha,\beta}\left(\boldsymbol{s},\boldsymbol{t},\pa\right)L(\boldsymbol{s},\boldsymbol{t},\pa),\\
&R_{\gamma}( \boldsymbol{s} +\boldsymbol{e}_{\alpha}-e_{\beta},\boldsymbol{t},\pa)P_{\alpha,\beta}\left(\boldsymbol{s},\boldsymbol{t},\pa\right)=
P_{\alpha,\beta}\left(\boldsymbol{s},\boldsymbol{t},\pa\right)R_{\gamma}(\boldsymbol{s},\boldsymbol{t},\pa),\\
&\frac{\pa P_{\alpha,\beta}\left(\boldsymbol{s},\boldsymbol{t},\pa\right)}{\pa t_{\gamma j}}= B_{\gamma j} (\boldsymbol{s}+\boldsymbol{e}_{\alpha}-\boldsymbol{e}_{\beta},\boldsymbol{t},\pa)
P_{\alpha,\beta}\left(\boldsymbol{s},\boldsymbol{t},\pa\right)-P_{\alpha,\beta}\left(\boldsymbol{s},\boldsymbol{t},\pa\right)
B_{\gamma j}(\boldsymbol{s},\boldsymbol{t},\pa).
\end{split}
\end{equation}These are the Lax equations of the multicomponent KP hierarchy.

For a fixed $\bs{s}$,  the first two equations of \eqref{eq10_28_3} are the Lax equations of   the multicomponent KP hierarchy proposed in \cite{2}. The other three equations in \eqref{eq10_28_3} which determine the variations of $L, R_1,\ldots, R_N$  with respect to the charge variable $\boldsymbol{s}$ are analogous to those proposed in the work on modified KP hierarchy \cite{3}. Therefore, the bilinear relation formulation of the multicomponent KP hierarchy \eqref{eq12_24_1} contains more information than the Lax formulation proposed in \cite{2}. It is essentially a multicomponent \emph{modified} KP hierarchy. 

\vspace{0.5cm}
\noindent
\textbf{Acknowledgment} We are grateful to K. Takasaki and T. Takebe for the useful discussion and helpful comments.


\begin{thebibliography}{10}
\bibitem{6}
E. Date, M. Kashiwara, M. Jimbo, and T. Miwa,
  \emph{Transformation groups for soliton equations}, Nonlinear integrable
  systems---classical theory and quantum theory (Kyoto, 1981), World Sci.
  Publishing, Singapore, 1983, pp.~39--119.
  
\bibitem{2}E. Date, M. Jimbo, M. Kashiwara and T. Miwa, \emph{Transformation groups for soliton equations III}, J. Phys. Soc. Japan \textbf{50} (1981), 3806--3812.

\bibitem{5} L. Dickey, \emph{Soliton equations and hamiltonian systems}, World Scientific, Singapore, 1991.


  
\bibitem{4}V. Kac, J. van de Leur, \emph{The $n$-component KP hierarchy and representation theory}, in: A.S. Fokas, V.E. Zakharov (Eds.), \emph{Important developments in soliton theory}, Springer-Verlag, Berlin, Heidelberg, 1993.


\bibitem{8} M. Sato, \emph{Soliton equations as dynamical systems on infinite dimensional Grassmann manifolds}, RIMS Kokyuroku (1981), 30.

\bibitem{7}   M. Sato, \emph{The KP hierarchy and infinite-dimensional Grassmann manifolds}, in \emph{Theta functions--Bowdoin 1987, Part 1},
Proc. Sympos. Pure Math., 49, Part 1, Amer. Math. Soc., Providence,
RI, (1989), pp. 51–-66.

 
\bibitem{1} K. Takasaki and T. Takebe, \emph{Universal Whitham hierarchy, dispersionless Hirota equations and
multicomponent KP hierarchy}, Physica D \textbf{235} (2007), 109-125.

\bibitem{9}
K. Takasaki and T. Takebe, \emph{Integrable hierarchies
and
  dispersionless limit}, Rev. Math. Phys. \textbf{7} (1995),   743--808.
  
\bibitem{10} K. Takasaki, \emph{Differential Fay identities and auxiliary linear problem of integrable hierarchies}, arXiv:0710.5356.  
  
\bibitem{3} T. Takebe, \emph{A Note on the modi¢ed KP hierarchy and its
(yet another) dispersionless limit}, Lett. Math. Phys. \textbf{59} (2002), 157-172.


\end{thebibliography}
\end{document}